\documentclass[11pt]{article}
\usepackage{}
\usepackage{acronym}
\usepackage{amssymb}
\usepackage{amsfonts}
\usepackage{mathrsfs}
\usepackage{graphicx}
\usepackage{amsmath}
\usepackage{color}
\usepackage{amsthm}
\usepackage{epstopdf}
\usepackage{braket}
\usepackage{pifont,bm}
\usepackage{tikz}
\usepackage{enumerate}
\usepackage{cite}
\usepackage{geometry}
\usepackage{hyperref}
\usepackage{mathtools}
\usepackage{setspace}
\usepackage{stmaryrd}
\usepackage{subcaption}
\usepackage{epsfig}
\usepackage{setspace}
\usepackage{booktabs}
\usepackage{threeparttable}
\usepackage{diagbox}

\theoremstyle{definition}

\makeatletter
\def\@biblabel#1{[#1]}
\makeatother

\makeatletter \@addtoreset{equation}{section}

\begin{document}

\begin{titlepage}
\title{\bf{$PT$ Symmetric PINN for integrable nonlocal  equations: Forward and inverse problems 
\footnote{
Corresponding authors.\protect\\
\hspace*{3ex} E-mail addresses: ychen@sei.ecnu.edu.cn (Y. Chen)}
}}
\author{Wei-Qi Peng, Yong Chen$^{a,b,*}$\\
\small \emph{$^{a}$School of Mathematical Sciences, Shanghai Key Laboratory of PMMP} \\
\small \emph{East China Normal University, Shanghai, 200241, China} \\
\small \emph{$^{b}$College of Mathematics and Systems Science, Shandong University }\\
\small \emph{of Science and Technology, Qingdao, 266590, China} \\
\date{}}
\thispagestyle{empty}
\end{titlepage}
\maketitle

\vspace{-0.5cm}
\begin{center}
\rule{15cm}{1pt}\vspace{0.3cm}

\parbox{15cm}{\small
{\bf Abstract}\\
\hspace{0.5cm}
Since the $PT$-symmetric nonlocal equations contain the physical information of the $PT$-symmetric, it is very appropriate to embed the physical information of the $PT$-symmetric into the loss function of PINN, named PTS-PINN. For general $PT$-symmetric nonlocal equations, especially those equations involving the derivation of nonlocal terms, due to the existence of nonlocal terms, directly using the original PINN method to solve such nonlocal equations will face certain challenges. This problem can be solved by the PTS-PINN method which can be illustrated in two aspects. First, we treat the nonlocal term of the equation as a new local component, so that the equation is coupled at this time. In this way, we successfully avoid differentiating nonlocal terms in neural networks. On the other hand, in order to improve the accuracy, we make a second improvement, which is to embed the physical information of the $PT$-symmetric into the loss function. Through a series of independent numerical experiments, we evaluate the efficacy of PTS-PINN in tackling the forward and inverse  problems  for the nonlocal nonlinear Schr\"{o}dinger (NLS) equation, the nonlocal derivative NLS equation, the nonlocal (2+1)-dimensional NLS equation, and the nonlocal three wave interaction systems. The numerical experiments demonstrate that PTS-PINN has good performance. In particular, PTS-PINN has also demonstrated an extraordinary ability to learn large space-time scale rogue waves for nonlocal equations.}

\vspace{0.5cm}
\parbox{15cm}{\small{

\vspace{0.3cm} \emph{Key words:}  $PT$-symmetric nonlocal equations; PTS-PINN; Rogue waves; Forward and inverse problems.\\

\emph{PACS numbers:}  02.30.Ik, 05.45.Yv, 04.20.Jb. } }
\end{center}
\vspace{0.3cm} \rule{15cm}{1pt} \vspace{0.2cm}

\section{Introduction}
Integrable nonlinear wave equations are widely acknowledged for their pivotal role in the realm of mathematical physics and engineering, encompassing fields such as fluid mechanics, plasma physics, ocean communication, and more. Within the domain of soliton theories, the pursuit of exact solutions for integrable equations has become a prominent subject, attracting an increasing number of researchers. Over the past decade, researchers have proposed numerous methods to tackle the challenge of solving integrable equations. Notable approaches include the Hirota bilinear method \cite{Peng-MMAS1}, Darboux transformation (DT) \cite{Peng-MMAS2}, inverse scattering transformation (IST) \cite{Peng-MMAS3}, and the physics informed neural networks (PINNs) algorithm \cite{Peng-MMAS4}, among others. Notably, significant attention has recently been directed towards the investigation of integrable nonlocal equations. It is particularly intriguing to explore the application of the PINNs algorithm in solving nonlocal integrable equations.

The first integrable nonlocal equation was the nonlocal nonlinear Schr\"{o}dinger (NLS) equation proposed by Ablowitz et al. \cite{Peng-MMAS44}, denoted as
\begin{align}\label{0.1}
iq_{t}(x,t)+q_{xx}(x,t)+\frac{1}{2}q^{2}(x,t)q^{\ast}(-x,t)=0,
\end{align}
where the asterisk $\ast$ signifies complex conjugation and $q^{\ast}(-x, t)$ represents a nonlocal term. Notably, the nonlocal NLS equation \eqref{0.1} incorporates the $PT$-symmetric potential $V(x,t)=q(x, t)q^{\ast}(-x, t)$, demonstrating invariance under the $x\rightarrow -x$ transformation and complex conjugation. Consequently, if $q(x, t)$ is a solution, then $q^{\ast}(-x, t)$ is also a solution. $PT$-symmetry, initially introduced by Bender and colleagues in quantum mechanics, posits that the non-Hermitian Hamiltonian manifests a wholly genuine spectrum \cite{Peng-MMAS41}. It is noteworthy that $PT$ symmetric systems have garnered considerable attention in optics and various other physical domains in recent years \cite{Yangbo25,Yangbo26}. Ref. \cite{Yangbo27} reported the application of the $PT$-symmetric NLS equation in unconventional magnetic systems. In Ref.\cite{JMP-4}, the IST method was proposed to study the Cauchy problem of the nonlocal NLS equation \eqref{0.1}, and soliton solutions were given. General higher-order solitons, freak wave and rational solutions of the nonlocal NLS were derived in \cite{Yangchen1,Yangchen2,Yangchen3}. The nondegenerate one- and two-soliton solutions of the nonlocal NLS equation were obtained by using the nonstandard Hirota method \cite{Yangchen4}. Subsequent to the introduction of the nonlocal $PT$-symmetric NLS equation, a rapid influx of new nonlocal integrable equations has been documented.

Rogue waves, also known as killer waves, have garnered significant attention in recent years in both maritime environments and optical fibers \cite{YangLMP2,YangLMP3}. Both theoretical predictions and experimental observations of rogue waves have additionally been documented \cite{YangLMP31,YangLMP32}. The first-order rational solution of the NLS proposed by
Peregrine was instrumental in explicating the phenomenon of rogue waves \cite{YangLMP4}. Subsequent to this, analytical rogue wave solutions have been derived for numerous integrable systems, such as the AB system \cite{YangLMP13}, the three wave
interaction equation \cite{YangLMP15}, the Kundu-Eckhaus(KE) equation \cite{YangLMP16}, and many others.
The majority of previously derived rogue waves pertained to local integrable equations. In contrast,
rogue waves in nonlocal integrable equations are an intriguing and newfangled research area and have been generated for various nonlocal integrable equations, including nonlocal $PT$-symmetric NLS equation \cite{YangboNLS}, nonlocal derivative
NLS equation \cite{DNLS2}, nonlocal Davey-Stewartson (DS) systems \cite{YangboDS, HeDS} and so on.

In this article, we study data-driven solutions, especially the rogue wave solutions,  in several integrable nonlocal equations via using
PINN deep learning method. As typically concrete examples, we focus on the nonlocal NLS equation \eqref{0.1}, the nonlocal derivative NLS equation \cite{Yangbo17}
\begin{align}\label{jia1}
iq_{t}(x,t)-q_{xx}(x,t)-(q^{2}(x,t)q^{\ast}(-x,t))_{x}=0,
\end{align}
the nonlocal (2+1)-dimensional NLS equation \cite{Peng-MMAS52,Peng-MMAS}
\begin{equation}
\begin{aligned}\label{jia2}
&iq_{t}+q_{xy}+qr=0,\\
&r_{y}=[q(x,y,t)q(-x,-y,t)^{\ast}]_{x},
\end{aligned}
\end{equation}
and the nonlocal three wave interaction systems \cite{Ab-siam}
\begin{equation}
\begin{aligned}\label{jia3}
&q_{1t}+aq_{1x}-q_{2}(-x,-t)q_{3}(-x,-t)=0,\\
&q_{2t}+bq_{2x}-q_{1}(-x,-t)q_{3}(-x,-t)=0,\\
&q_{3t}+cq_{3x}+q_{1}(-x,-t)q_{2}(-x,-t)=0.
\end{aligned}
\end{equation}
For the nonlocal derivative NLS equation, soliton solutions have been previously presented via the IST \cite{DNLS1}. Additionally, investigations into its periodic wave solutions and rogue waves on a double-periodic background have been conducted using the DT \cite{Yangbo17,DNLS2}.
In the context of the nonlocal (2+1)-dimensional NLS equation, higher-order rational solutions and interaction solutions have been derived through a generalized DT \cite{Wangzhen}. Furthermore, employing the long wave limit method, the analysis of rogue waves and semi-rational solutions for the nonlocal (2+1)-dimensional NLS equation has been explored \cite{Peng-MMAS52,Peng-MMAS}. For the nonlocal three wave interaction systems, the $a, b, c$ are constant parameters. The inverse scattering transformation of the three wave interaction equation is studied in detail and its soliton solution is derived \cite{Ab-siam}.

In the context of the burgeoning computational capabilities of computers, deep learning has emerged as a highly effective technical tool for solving partial differential equations (PDEs). Notably, benefiting from the general approximation theorem \cite{Miu14} and automatic differentiation techniques \cite{PuChen420}, Raissi et al. have proposed a deep learning framework based on physical constraints to efficiently solve PDEs, which they call physical information neural networks (PINN) \cite{Peng-MMAS4}. The fundamental idea of underlying PINN involves embedding PDEs into the loss function of a neural network, enabling the numerical solution of PDEs through gradient descent with respect to the loss function, along with the incorporation of initial and boundary conditions. Compared to traditional numerical methods, PINN demonstrate the capability to solve PDEs with limited data while also facilitating the discovery of equation parameters. Subsequently, PINN have garnered considerable attention across various fields. In the realm of integrable systems, Professor Chen's team pioneered the utilization of the PINN method to construct data-driven soliton solutions, breath waves, rogue waves, and rogue periodic waves for diverse nonlinear evolution equations, including the KdV equation, derivative NLS equation, Chen-Lee-Liu equation, and variable coefficient model, among others \cite{Peng-PD29,Peng-PD30,Peng-PD31,Peng-PD32,Peng-PD,Zhujy}. Concurrently, other scholars have presented noteworthy results for data-driven solutions of equations such as the Gross-Pitaevskii equation, Boussinesq equation, NLS equation, coupled NLS equation, Benjamin-Ono equation, saturable NLS equation with $PT$-symmetric
potentials and so on\cite{Peng-PD33,Peng-PD34,Peng-PD35,Peng-PD36,Peng-PD37,Peng-PD38,Peng-PD39,jia1,jia2,jia3,jia4,DS-Chaos}. Despite the notable successes achieved thus far, PINN face challenges when addressing more intricate problems that demand increased precision and efficiency. Therefore, an enhanced version of PINN has been developed. Noteworthy improvements include the embedding of gradient information from residual PDEs into the loss function, leading to the introduction of a gradient-enhanced PINN \cite{Zhang-JCP7}. Additionally, Lin and Chen introduced a two-stage PINN by incorporating conserved quantities into mean square error losses \cite{Zhang-JCP8}. Techniques such as adaptive activation functions \cite{Miu33} and self-adaptive loss balanced PINN \cite{Miu51} have been introduced to expedite network training and improve accuracy. The development of symmetric enhanced physical information neural networks further enhances accuracy by incorporating Lie's symmetry or the nonclassical symmetry of PDEs into the loss function \cite{Zhang-JCP}.

Nonlocal integrable equations distinguish themselves from local equations by featuring unique spatial and/or temporal couplings, leading to novel physical effects and inspiring innovative applications. Notably, the nonlocal integrable equations has $PT$-symmetry property, which can be embedded into the loss function to improve the precision of PINN algorithm in solving the nonlocal integrable equations. Consequently, the application of PINN becomes imperative for studying such nonlocal integrable equations. However, owing to the presence of nonlocal terms, employing the original PINN method directly for addressing such nonlocal equations poses specific challenges. This paper introduces PTS-PINN to solve the general $PT$-symmetric nonlocal equations, particularly nonlocal equations with derivatives of nonlocal terms. The fundamental concepts of PTS-PINN encompass two key aspects. Firstly, we consider the nonlocal term of the equation as a novel local component, resulting in the coupling of the equation. This approach effectively eliminates the need to differentiate nonlocal terms within neural networks. Secondly, to enhance accuracy, we implement a second improvement by incorporating the physical information of $PT$-symmetry into the loss function

The outline of this paper is organized as follows:  In Section 2, we  briefly give a description for the main ideas of the PTS-PINN. In Section 3, we give out several sets of independent experiments for the nonlocal NLS equation, the nonlocal derivative NLS equation,
the nonlocal (2+1)-dimensional NLS equation to illustrate the effectiveness of PTS-PINN. The data-driven rogue wave solution, periodic wave solution, breather wave solution are investigated via using PTS-PINN scheme. Moreover, applying the PTS-PINN method, Section 4 discusses the data-driven parameters discovery for the nonlocal (2+1)-dimensional NLS equation and nonlocal three wave interaction systems. Conclusion is given out in last section.

\section{Main ideas of the $PT$-symmetry PINN}

This section briefly introduces the utilization of $PT$-symmetry PINN for addressing nonlocal complex PDEs. Consider a general nonlocal complex PDE with the following form:
\begin{align}\label{1}
\left\{
\begin{array}{lr}
iq(\textbf{x},t)_{t}+\mathcal{N}[q(\textbf{x},t),q(-\textbf{x},t)]=0,\quad \textbf{x}\in \Omega, \quad t\in [T_{0},T_{1}],\\
q(\textbf{x},T_{0})=h_{0}(\textbf{x}),\quad \textbf{x}\in\Omega,\\
q(\textbf{x},t)=h_{\Gamma}(\textbf{x},t),\quad \textbf{x}\in\partial \Omega,\quad t\in [T_{0},T_{1}],
\end{array}
\right.
\end{align}
where the variable $\textbf{x}$ represents an $N$-dimensional vector denoted as $\textbf{x} = (x_{1}, x_{2}, \cdots , x_{N})$, and $\textbf{x}$
belongs to the spatial domain $\Omega$ and $t$ belongs to the temporal interval $[T_{0},T_{1}]$
 $\mathcal{N}[\cdot]$ represents a nonlinear differential operator,  signifying a  smooth function of $q(\textbf{x},t), q(-\textbf{x},t)$, and their $\textbf{x}$-derivatives
up to $r$-th order.  The boundary of the spatial domain $\Omega$ is denoted by $\partial\Omega$, the
 final  two equations in \eqref{1} correspond to the initial value condition and the Dirichlet boundary condition, respectively.
$q=q(\textbf{x},t)$ represents  a complex valued solution of the equation. To facilitate the separation of the real and
imaginary parts of Eq.\eqref{1}, we express $q(\textbf{x},t)$ as $u_{R}(\textbf{x},t)+iu_{I}(\textbf{x},t)$, where
$u_{R}$ and $u_{I}$ are the real and imaginary parts, respectively.

It is a widely acknowledged fact that a nonlocal equation featuring $PT$-symmetry is described by a $PT$ potential $V(\textbf{x},t)=q(\textbf{x},t)q^{\ast}(-\textbf{x},t)$. In the context of optics, the real component of  $V(\textbf{x},t)$,
denoted as $V_{R}(\textbf{x},t)$, serves as the  refractive-index profile, while the imaginary component of
$V(\textbf{x},t)$  signifies the
gain/loss distribution and is defined as  $V_{I}(\textbf{x},t)$. Based on
$V_{R}(\textbf{x},t)=V_{R}(-\textbf{x},t)$ and $V_{I}(\textbf{x},t)=-V_{I}(-\textbf{x},t)$, a $PT$ symmetric system can be designed. In order to apply $PT$-symmetry properties to the PINN loss function, we let $q(-\textbf{x},t)=v_{R}(\textbf{x},t)+iv_{I}(\textbf{x},t)$. Then we have
\begin{align}\label{1.1}
u_{R}(\textbf{x},t)=v_{R}(-\textbf{x},t),  \qquad u_{I}(\textbf{x},t)=v_{I}(-\textbf{x},t).
\end{align}
Substituting $q(\textbf{x},t)=u_{R}(\textbf{x},t)+iu_{I}(\textbf{x},t)$ and $q(-\textbf{x},t)=v_{R}(\textbf{x},t)+iv_{I}(\textbf{x},t)$ into Eq.\eqref{1},  we get two real equations
\begin{align}\label{2}
\left\{
\begin{array}{lr}
(u_{R})_{t}+\mathcal{N}_{R}[u_{R},u_{I},v_{R},v_{I}]=0,\\
(u_{I})_{t}+\mathcal{N}_{I}[u_{R},u_{I},v_{R},v_{I}]=0.
\end{array}
\right.
\end{align}
Extending the pioneering work of  PINN, we define a neural network approximation of the solution  $(u_{R}(\textbf{x},t), u_{I}(\textbf{x},t), v_{R}(\textbf{x},t), v_{I}(\textbf{x},t))$ as $(\hat{u}_{R}(\textbf{x},t; \theta), \hat{u}_{I}(\textbf{x},t; \theta), \hat{v}_{R}(\textbf{x},t; \theta), \hat{v}_{I}(\textbf{x},t; \theta))$, where $\theta$
 represents a set of network parameters. In particular, we consider a neural network with $M$ layers, encompassing one
input layer, $(M-1)$ hidden layers, and one output layer. The $m$th (m = 1, 2,\ldots, M-1) layer comprises $N_{m}$ neurons, signifying the transmission of an $N_{m}$-dimensional output vector to the $(m+1)$th layer as the input data. The interconnection  between two layers is
 established through the linear transformation $\mathcal{T}_{m}$ and the nonlinear activation function $\sigma(\cdot)$
\begin{align}\label{3}
\textbf{X}^{[m]}=\mathcal{T}_{m}(\textbf{X}^{[m-1]})=\sigma(\textbf{W}^{[m]}\textbf{X}^{[m-1]}+\textbf{b}^{[m]}),
\end{align}
where $\textbf{X}^{[m]}$
represent the state vector of the
$m$th layer node, with particular emphasis on the data $\textbf{X}^{[0]}$, which is the transpose of vector $(x_{1},x_{2},\ldots, x_{N},t)$. The weight matrix and bias vector of the $m$th layer are denoted by $\textbf{W}^{[m]}\in \mathbb{R}^{N_{m}\times N_{m-1}}$ and  $\textbf{b}^{[m]}\in \mathbb{R}^{N_{m}}$, respectively. The parameter set $\theta=\left\{\textbf{W}^{[m]}, \textbf{b}^{[m]}\right\}_{1\leq m\leq M}$ encompasses all weight matrices and bias vectors involved in the layers.  The hyperbolic tangent (tanh) function stands out as the prevailing activation function in neural networks employed for solving PDEs. Simultaneously, the Xavier initialization method is employed to initialize the weight matrix and bias vector. To articulate the PDEs residual, one can substitute $\hat{u}_{R}(\textbf{x},t; \theta), \hat{u}_{I}(\textbf{x},t; \theta), \hat{v}_{R}(\textbf{x},t; \theta), \hat{v}_{I}(\textbf{x},t; \theta)$ into Eq.\eqref{2}:
\begin{align}\label{4}
\left\{
\begin{array}{lr}
f_{A}(\textbf{x},t; \theta):=\frac{\partial}{\partial t}\hat{u}_{R}(\textbf{x},t; \theta)+\mathcal{N}_{R}[\hat{u}_{R}(\textbf{x},t; \theta), \hat{u}_{I}(\textbf{x},t; \theta), \hat{v}_{R}(\textbf{x},t; \theta), \hat{v}_{I}(\textbf{x},t; \theta)],\\
f_{B}(\textbf{x},t; \theta):=\frac{\partial}{\partial t}\hat{u}_{I}(\textbf{x},t; \theta)+\mathcal{N}_{I}[\hat{u}_{R}(\textbf{x},t; \theta), \hat{u}_{I}(\textbf{x},t; \theta), \hat{v}_{R}(\textbf{x},t; \theta), \hat{v}_{I}(\textbf{x},t; \theta)].
\end{array}
\right.
\end{align}
Utilizing automatic differentiation mechanism on $\hat{u}_{R}(\textbf{x},t; \theta), \hat{u}_{I}(\textbf{x},t; \theta), \hat{v}_{R}(\textbf{x},t; \theta), \hat{v}_{I}(\textbf{x},t; \theta)$, we obtain the residuals $f_{A}(\textbf{x},t; \theta), f_{B}(\textbf{x},t; \theta)$ for the PINN. However, it is crucial to note that the PDEs residual \eqref{4} does not encompass additional inherent physical properties, such as the $PT$-symmetry detailed in \eqref{1.1}, which are not explicitly considered during the neural network training process. Consequently, it is justifiable to incorporate the \eqref{1.1} into the loss function of PINN, aiming to further enhance the accuracy of the neural network.  This leads to the formulation of the other two residuals:
\begin{align}\label{4.1}
\left\{
\begin{array}{lr}
f_{R}(\textbf{x},t; \theta):=\hat{u}_{R}(\textbf{x},t; \theta)-\hat{v}_{R}(-\textbf{x},t; \theta),\\
f_{I}(\textbf{x},t; \theta):=\hat{u}_{I}(\textbf{x},t; \theta)-\hat{v}_{I}(-\textbf{x},t; \theta).
\end{array}
\right.
\end{align}
Subsequently, a multi-hidden-layer deep neural network is employed to train the network parameters associated with the latent functions $\hat{u}_{R}(\textbf{x},t; \theta), \hat{u}_{I}(\textbf{x},t; \theta), \hat{v}_{R}(\textbf{x},t; \theta), \hat{v}_{I}(\textbf{x},t; \theta)$ and  residual networks $f_{A}$, $f_{B}, f_{R}, f_{I}$. In order to achieve optimal training results, we formulate the following loss functions, which are subject to minimization using the L-BFGS optimization method \cite{wangya44}
\begin{align}\label{5}
Loss_{\theta}=&Loss_{\hat{u}_{R}}+Loss_{\hat{u}_{I}}+Loss_{\hat{v}_{R}}+Loss_{\hat{v}_{I}}\notag\\
&+Loss_{f_{A}}+Loss_{f_{B}}+Loss_{f_{R}}+Loss_{f_{I}},
\end{align}
with
\begin{align}\label{6}
\left\{
\begin{array}{lr}
Loss_{\hat{u}_{R}}=\frac{1}{N_{IB}}\sum_{i=1}^{N_{IB}}|\hat{u}_{R}(\textbf{x}_{IB}^{i},t_{IB}^{i}; \theta)-u_{R}^{i}|^{2},\\
Loss_{\hat{u}_{I}}=\frac{1}{N_{IB}}\sum_{i=1}^{N_{IB}}|\hat{u}_{I}(\textbf{x}_{IB}^{i},t_{IB}^{i}; \theta)-u_{I}^{i}|^{2},\\
Loss_{\hat{v}_{R}}=\frac{1}{N_{IB}}\sum_{i=1}^{N_{IB}}|\hat{v}_{R}(\textbf{x}_{IB}^{i},t_{IB}^{i}; \theta)-v_{R}^{i}|^{2},\\
Loss_{\hat{v}_{I}}=\frac{1}{N_{IB}}\sum_{i=1}^{N_{IB}}|\hat{v}_{I}(\textbf{x}_{IB}^{i},t_{IB}^{i}; \theta)-v_{I}^{i}|^{2},\\
Loss_{f_{A}}=\frac{1}{N_{f}}\sum_{j=1}^{N_{f}}|f_{\hat{u}}(\textbf{x}_{f}^{j},t_{f}^{j})|^{2},\\
Loss_{f_{B}}=\frac{1}{N_{f}}\sum_{j=1}^{N_{f}}|f_{\hat{v}}(\textbf{x}_{f}^{j},t_{f}^{j})|^{2},\\
Loss_{f_{R}}=\frac{1}{N_{f}}\sum_{j=1}^{N_{f}}|f_{R}(\textbf{x}_{f}^{j},t_{f}^{j})|^{2},\\
Loss_{f_{I}}=\frac{1}{N_{f}}\sum_{j=1}^{N_{f}}|f_{I}(\textbf{x}_{f}^{j},t_{f}^{j})|^{2},
\end{array}
\right.
\end{align}
where $\{\textbf{x}_{IB}^{i}, t_{IB}^{i}, u_{R}^{i}, u_{I}^{i}, v_{R}^{i}, v_{I}^{i}\}_{i=1}^{N_{IB}}$ represent the sampling initial boundary (IB) value training data points. Additionally,  $\{\textbf{x}_{f}^{j}, t_{f}^{j}\}_{j=1}^{N_{f}}$ denote the collocation points sampled for $f_{A}, f_{B}, f_{R}, f_{I}$.

Conversely, PINN also demonstrate efficacy in addressing the inverse problem associated with nonlocal integrable PDEs, where both undetermined parameters and numerical solutions are concurrently learned. In instances  there are unknown parameters in Eq.\eqref{1},these parameters can be acquired through the incorporation of additional measurements of $(\hat{u}_{R}(\textbf{x},t; \theta), \hat{u}_{I}(\textbf{x},t; \theta), \hat{v}_{R}(\textbf{x},t; \theta), \hat{v}_{I}(\textbf{x},t; \theta))$ into the set of labeled training points
 $\{\textbf{x}_{l}^{i}, t_{l}^{i}, u_{R}^{i}, u_{I}^{i}, v_{R}^{i}, v_{I}^{i}\}_{i=1}^{N_{l}}$.
This involves introducing an extra data loss term
\begin{align}\label{7}
Loss_{l}=&\frac{1}{N_{l}}\sum_{i=1}^{N_{l}}\left[|\hat{u}_{R}(\textbf{x}_{l}^{i},t_{l}^{i}; \theta)-u_{R}^{i}|^{2}+|\hat{u}_{I}(\textbf{x}_{l}^{i},t_{l}^{i}; \theta)-u_{I}^{i}|^{2}\right.\notag\\
&\left.+|\hat{v}_{R}(\textbf{x}_{l}^{i},t_{l}^{i}; \theta)-v_{R}^{i}|^{2}+|\hat{v}_{I}(\textbf{x}_{l}^{i},t_{l}^{i}; \theta)-v_{I}^{i}|^{2}\right],
\end{align}
thus the loss function for the inverse problem of PDEs is given by
\begin{align}\label{8}
Loss_{\theta}=&Loss_{l}+Loss_{\hat{u}_{R}}+Loss_{\hat{u}_{I}}+Loss_{\hat{v}_{R}}+Loss_{\hat{v}_{I}}\notag\\
&+Loss_{f_{A}}+Loss_{f_{B}}+Loss_{f_{R}}+Loss_{f_{I}}.
\end{align}
For convenience, the IB value training points also can be regarded as part of the labeled training points. Therefore, the final loss function can be  written as
\begin{align}\label{9}
Loss_{\theta}=Loss_{l}+Loss_{f_{A}}+Loss_{f_{B}}+Loss_{f_{R}}+Loss_{f_{I}}.
\end{align}

In contrast to the original PINN, the optimization objective in PTS-PINN involves minimizing the loss function, which now includes the $PT$-symmetry residual in addition to the PINN loss function. This optimization process entails updating the weights and biases. To provide a comprehensive overview of the PTS-PINN methodology, we present a schematic diagram in Fig. 1.\\
{\centerline{\includegraphics[width=14.0cm,height=6.0cm,angle=0]{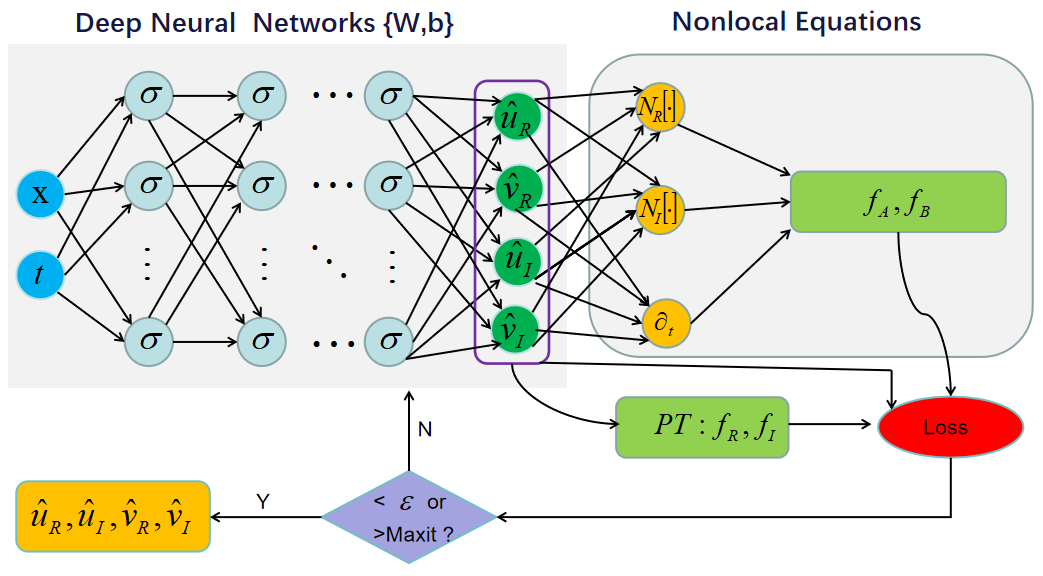}}}\\
{\centerline\noindent { \small \textbf{Figure 1.} (Color online) The PTS-PINN scheme solving the nonlocal equation.}}\\

\section{Numerical experiments for forward problems}
In this section, we  mainly consider the positive problems of several classes nonlocal equations, including nonlocal NLS equation, the nonlocal derivative NLS equation, and the nonlocal (2+1)-dimensional NLS equation.  Here, we use Tensorflow to implement neural networks and all numerical experiments shown here are run on a DELL Precision 7920 Tower computer with 2.10 GHz 8-core Xeon Silver 4110 processor and 64-GB memory.
The accuracy of the estimation is assessed using the relative $\mathbb{L}_{2}$ error, a metric that serves as a reliable indicator of algorithm effectiveness. The exact value $q(\textbf{x}_{k},t_{k})$ and
the trained approximation $\hat{q}(\textbf{x}_{k},t_{k})$ are inferred at the data points $\{ \textbf{x}_{k},t_{k}\}_{k=1}^{N}$, then the error can be defined as
\begin{align}\label{10}
\mbox{Error}=\frac{\sqrt{\sum_{k=1}^{N}\left\vert q(\textbf{x}_{k},t_{k})-\hat{q}(\textbf{x}_{k},t_{k}) \right\vert^{2}}}{\sqrt{\sum_{k=1}^{N}\left\vert q(\textbf{x}_{k},t_{k})\right\vert^{2}}}.
\end{align}

\subsection{Nonlocal NLS equation}
The nonlocal NLS equation along with Dirichlet boundary conditions reads as \cite{Liuwei}
\begin{align}\label{11}
\left\{
\begin{array}{lr}
iq_{t}(x,t)+q_{xx}(x,t)+\frac{1}{2}q^{2}(x,t)q^{\ast}(-x,t)=0,\qquad x\in [-5,5],\quad t \in [-5,5],\\
\\
q(x,-5)=e^{-\frac{5i}{2}}(1-\frac{4(1-5i)}{x^{2}+26}),\\
\\
q(-5,t)=q(5,t)=e^{\frac{it}{2}}(1-\frac{4it+4}{t^{2}+26}).
\end{array}
\right.
\end{align}
The analytic solution is $q(x,t)=e^{\frac{it}{2}}(1-\frac{4it+4}{t^{2}+x^{2}+1})$, which is the rogue wave solution. Our goal is to derive the rogue wave solution $q(x,t)$ using PTS-PINN, employing a 5-layer fully-connected neural network with 20 neurons per hidden layer and employing the hyperbolic tangent (tanh) activation function. In MATLAB, the traditional finite difference method is applied to generate the initial training data by discretizing the analytic solution across spatial regions
$[-5, 5]$ into 512 points and time regions $[-5, 5]$
 into 400 points. The original training data comprises both IB data and inner points. Using the Latin hypercube sampling (LHS) method \cite{Peng-PD60}, $N_{IB}=300$ data points are randomly extracted from the original IB data, and $N_{f}=10000$ points serve as collocation points from the inner points. Executing the PTS-PINNs scheme with this training data, the data-driven rogue wave solution
$q(x,t)$ is learned with an $\mathbb{L}_{2}$-norm error of 9.491509e-03 compared to the exact solution. The total learning process involves 30,149 iterations and takes approximately 968.9108 seconds. The corresponding dynamic behaviors are depicted in Fig. 2, illustrating the wave propagation pattern along the $x$-axis and density plots. In Fig. 2, the error range of -0.025 to 0.025 between the learned dynamics and true dynamics indicates the effectiveness of PTS-PINN. Furthermore, we conduct a analysis of the relative $\mathbb{L}_{2}$ errors from different number of hidden layers, neurons per layer,  IB points and collocation points in Table 1, Table 2, respectively. The results presented in these tables indicate a noticeable results: the $\mathbb{L}_{2}$ relative errors of PTS-PINN consistently achieve e-02 or even e-03. In summary, the superior performance of PTS-PINN is evident.\\
{\centerline{\includegraphics[width=10.0cm,height=5.0cm,angle=0]{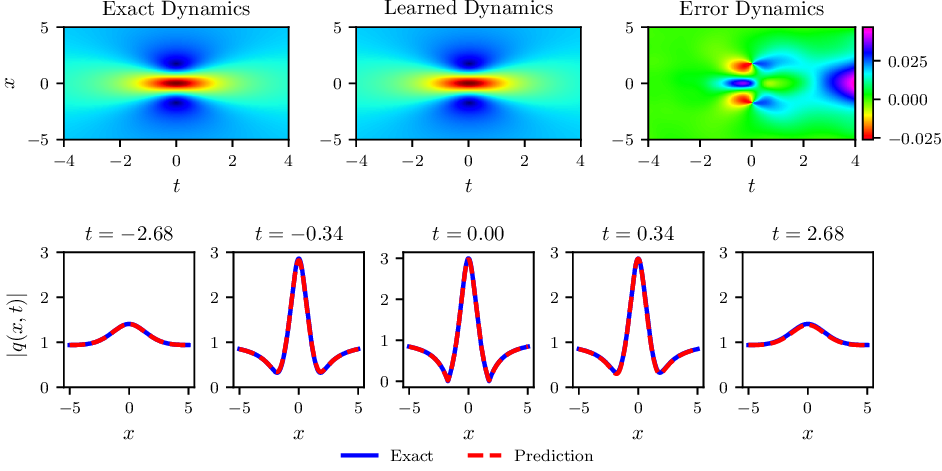}}}\\
\noindent { \small \textbf{Figure 2.} (Color online) Solving the data-driven rogue wave solution $q(x, t)$ for nonlocal NLS equation \eqref{11} via PTS-PINN:
The exact, learned and error dynamics density plots, and the sectional drawings which contain the learned and explicit rogue wave solution $q(x, t)$ at the  five distinct times $t=-3.35, t=-0.42, t=0, t=0.42, t=3.35$.}\\

\begin{table}[htbp]
  \caption{nonlocal NLS equation: $\mathbb{L}_{2}$ relative errors of PTS-PINN for different number of hidden layers and neurons, $N_{IB}=300, N_{f}=10000$}
  \label{Tabl}
  \centering
  \begin{tabular}{l|cccc}
  \toprule
  \diagbox{\textbf{Layers}}{\textbf{Neurons}} & 20 & 30 & 40\\
  \hline
  5   & 9.491509e-03 & 1.702184e-02 & 2.074911e-02\\
  6   & 1.581018e-02 & 2.064653e-02 & 1.014334e-02\\
  7   & 1.746864e-02 & 9.272919e-03 & 9.523377e-03\\
  \bottomrule
  \end{tabular}
\end{table}

\begin{table}[htbp]
  \caption{nonlocal NLS equation: $\mathbb{L}_{2}$ relative errors PTS-PINN for different number of  IB points and  collocation points, 5 layer  neural network and 20 neurons per layer}
  \label{Tab2}
  \centering
  \begin{tabular}{l|cccc}
  \toprule
  \diagbox{\textbf{IB points}}{\textbf{Collocation points}} & 5000 & 10000 & 20000\\
  \hline
  300   & 3.367488e-01 & 9.491509e-03 & 1.735005e-02\\
  400   & 4.115099e-02 & 1.912271e-02 & 2.051094e-02\\
  500   & 4.052362e-02 & 3.716130e-02 & 1.874488e-02\\
  \bottomrule
  \end{tabular}
\end{table}

\subsection{Nonlocal derivative NLS equation}
The nonlocal derivative NLS equation with Dirichlet boundary conditions \cite{Yangbo17}
\begin{align}\label{13}
\left\{
\begin{array}{lr}
iq_{t}(x,t)-q_{xx}(x,t)-(q^{2}(x,t)q^{\ast}(-x,t))_{x}=0,\qquad x\in [-5,5],\quad t \in [0,1],\\
\\
q(x,0)=\frac{(8-8i)(ie^{-3ix}-16e^{-ix})}{(ie^{-2ix}+16)^{2}},\\
\\
q(-5,t)=\frac{(8-8i)(ie^{i(t+15)}-16e^{i(t+5)})}{(ie^{10i}+16)^{2}},\ q(5,t)=\frac{(8-8i)(ie^{i(t-15)}-16e^{i(t-5)})}{(ie^{-10i}+16)^{2}}.
\end{array}
\right.
\end{align}
The analytic solution for the nonlocal derivative NLS equation \eqref{13} is given by $q(x,t)=\frac{(8-8i)(ie^{i(t-3x)}-16e^{i(t-x)})}{(ie^{-2ix}+16)^{2}}$, which represents a periodic wave solution.
The objective is to learn this solution
$q(x,t)$ using PTS-PINN. The neural network architecture consists of a 5-layer fully-connected network with 20 neurons per hidden layer, employing the hyperbolic tangent (tanh) activation function. The spatial region
$[-5, 5]$ is discretized into 512 points, and the time region $[0, 1]$
 into 100 points. A total of $N_{IB}=300$ points are randomly extracted from the IB data, while $N_{f}=10000$ points serve as collocation points within the inner region. In the context of the PTS-PINN scheme, these training data are processed to learn the data-driven periodic wave solution $q(x, t)$. The obtained solution demonstrates an $\mathbb{L}_{2}$-norm error of 6.460356e-04 in comparison to the exact solution. The entire learning process comprises 20,262 iterations and takes approximately 874.7039 seconds. The corresponding dynamic behavior is depicted in Fig. 3, illustrating that the error between the learned dynamics and the actual dynamics is negligible. Besides, we also discuss $\mathbb{L}_{2}$ relative errors of PTS-PINN for different number of hidden layers, neurons, IB points and collocation points in Table 3 and Table 4.\\
{\centerline{\includegraphics[width=10.0cm,height=5.0cm,angle=0]{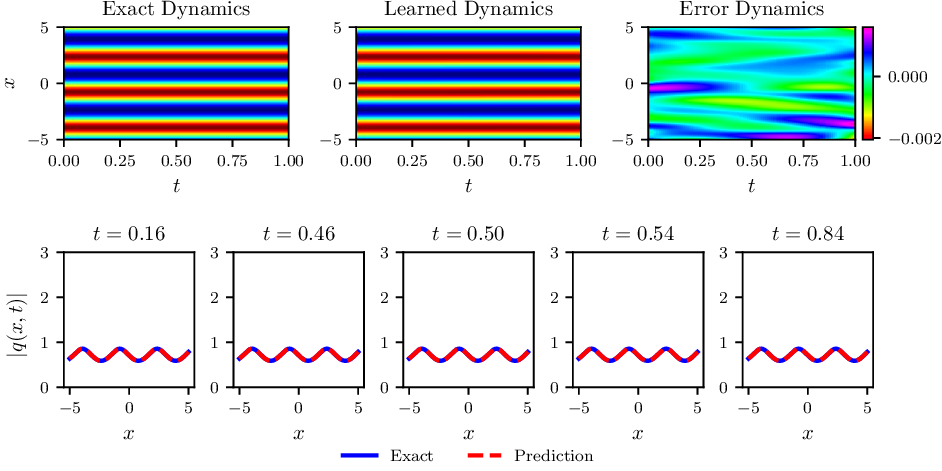}}}\\
\noindent { \small \textbf{Figure 3.} (Color online) Solving the data-driven periodic wave solution $q(x, t)$ for nonlocal derivative NLS equation \eqref{11} via PTS-PINN:
The exact, learned and error dynamics density plots, and the sectional drawings which contain the learned and explicit rogue wave solution $q(x, t)$ at the  five distinct times $t=0.16, t=0.46, t=0.50, t=0.54, t=0.84$.}\\

\begin{table}[htbp]
  \caption{nonlocal derivative NLS equation: $\mathbb{L}_{2}$ relative errors of PTS-PINN for different number of hidden layers and neurons, $N_{IB}=300, N_{f}=10000$}
  \label{Tab3}
  \centering
  \begin{tabular}{l|cccc}
  \toprule
  \diagbox{\textbf{Layers}}{\textbf{Neurons}} & 20 & 30 & 40\\
  \hline
  4   & 1.164461e-03 & 8.190090e-04 & 3.230923e-04\\
  5   & 6.460356e-04 & 3.642038e-04 & 4.218157e-04\\
  6   & 7.456470e-04 & 4.480699e-04 & 5.099400e-04\\
  \bottomrule
  \end{tabular}
\end{table}

\begin{table}[htbp]
  \caption{nonlocal derivative NLS equation: $\mathbb{L}_{2}$ relative errors of PTS-PINN for different number of IB
points and collocation points, 5 layer neural network and 20 neurons per layer}
  \label{Tab4}
  \centering
  \begin{tabular}{l|cccc}
  \toprule
  \diagbox{\textbf{IB points}}{\textbf{Collocation points}} & 5000 & 10000 & 15000\\
  \hline
  100   & 1.947399e-03 & 3.267939e-03 & 1.237726e-03\\
  200   & 1.574604e-03 & 1.340912e-03 & 1.889610e-03\\
  300   & 8.779340e-04 & 6.460356e-04 & 1.327981e-03\\
  \bottomrule
  \end{tabular}
\end{table}

\subsection{Nonlocal (2+1)-dimensional NLS equation}
The nonlocal (2+1)-dimensional NLS equation is as follows
\begin{equation}
\begin{aligned}\label{15}
&iq_{t}+q_{xy}+qr=0,\quad x\in [-12,12],\ y\in [-8,8],\ t \in [-4,4],\\
&r_{y}=[q(x,y,t)q(-x,-y,t)^{\ast}]_{x},
\end{aligned}
\end{equation}
and the analytic solution is \cite{Peng-MMAS52}
\begin{equation}
\begin{aligned}\label{16}
&q(x,y,t)=\frac{1+ie^{ix+iy-t}+ie^{-ix-iy-t}-2e^{-2t}}{1+e^{ix+iy-t}+e^{-ix-iy-t}+2e^{-2t}},\\
&r(x,y,t)=-\frac{2e^{-t}(2e^{-2t+ix+iy}+2e^{-2t-ix-iy}+e^{i(x+y)}+e^{-i(x+y)}+4e^{-t})}{(1+e^{ix+iy-t}+e^{-ix-iy-t}+2e^{-2t})^{2}},
\end{aligned}
\end{equation}
which is the breather wave solution. It needs to be emphasized that the $r(x,y,t)$ is a real function, which leads to a additional loss function, given by
\begin{align}\label{16.1}
Loss_{r}=\frac{1}{N_{IB}}\sum_{i=1}^{N_{IB}}|\hat{r}(x_{IB}^{i},y_{IB}^{i},t_{IB}^{i}; \theta)-r^{i}|^{2}.
\end{align}
We endeavor to acquire the solutions $q(x, y, t)$ and $r(x, y, t)$ through the employment of PTS-PINN, employing a 6-layer fully-connected neural network with 40 neurons per hidden layer and utilizing tanh activation. The original training data is obtained by the traditional finite difference method. This involves discretizing the $x$-spatial region $[-12, 12]$ into 64 points, the $x$-spatial region $[-8, 8]$ into 64 points, and the time region $[-4, 4]$ into 32 points. We randomly select $N_{IB}=3000$ points from the original IB data and designate $N_{f}=20000$  points as collocation points from the inner region. Processing these curated training data within the PTS-PINN scheme, we successfully learn the data-driven rogue wave solution $q(x, y, t), r(x, y, t)$. The achieved solution demonstrates an $\mathbb{L}_{2}$-norm error of 6.421177e-04 and 2.315614e-03 when compared to the exact solution. The total learning process involves 22,180 iterations and consumes approximately 1561.3419 seconds. The corresponding dynamic behaviors are presented in Fig. 4 and Fig. 5. The data presented in Table 5, and Table 6  outline the relative $\mathbb{L}_{2}$ error for PTS-PINN under various configurations, including different numbers of hidden layers, IB points, collocation points and neurons per layer. The consistent trend across these tables underscores the overall superior performance of PTS-PINN.\\

{\rotatebox{0}{\includegraphics[width=4.0cm,height=4.0cm,angle=0]{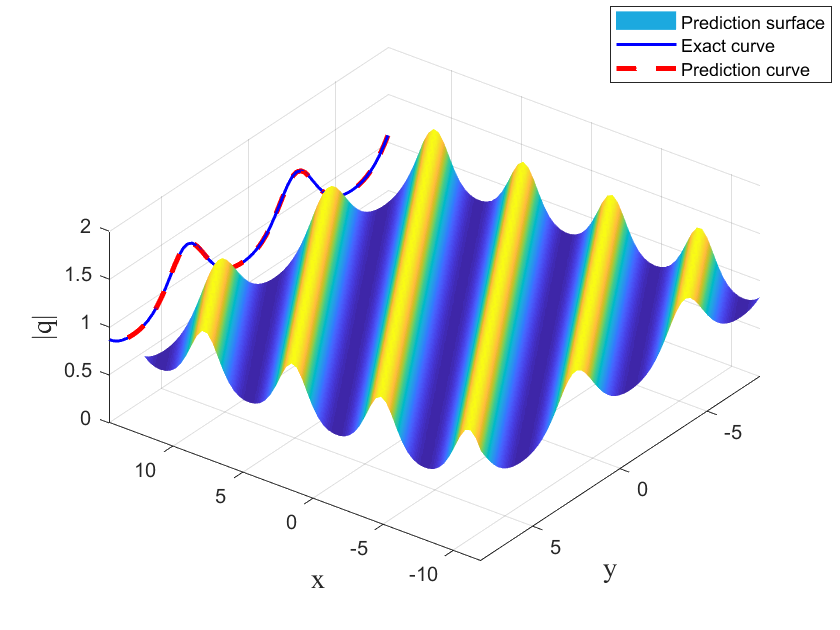}}}
~~~~
{\rotatebox{0}{\includegraphics[width=4.0cm,height=4.0cm,angle=0]{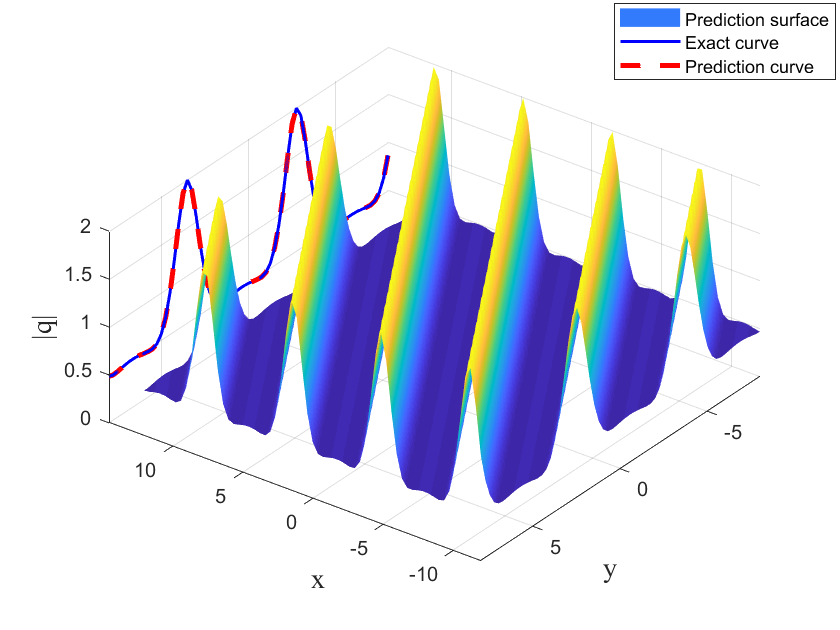}}}
~~~~
{\rotatebox{0}{\includegraphics[width=4.0cm,height=4.0cm,angle=0]{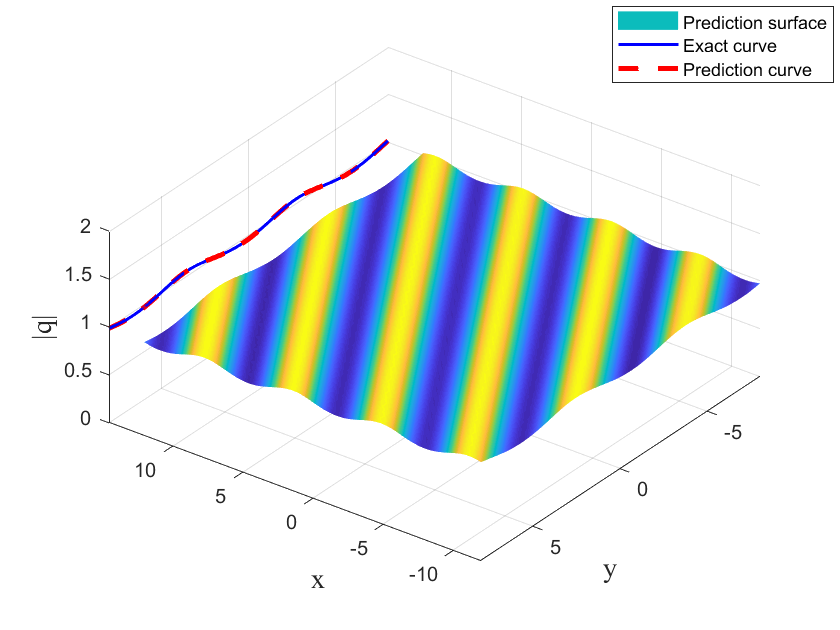}}}

$\quad\qquad\quad\quad\textbf{(a)}\quad\qquad\qquad\qquad\qquad\quad\quad\textbf{(b)}\qquad\qquad\qquad\quad\qquad\quad\textbf{(c)}$

{\rotatebox{0}{\includegraphics[width=4.0cm,height=4.0cm,angle=0]{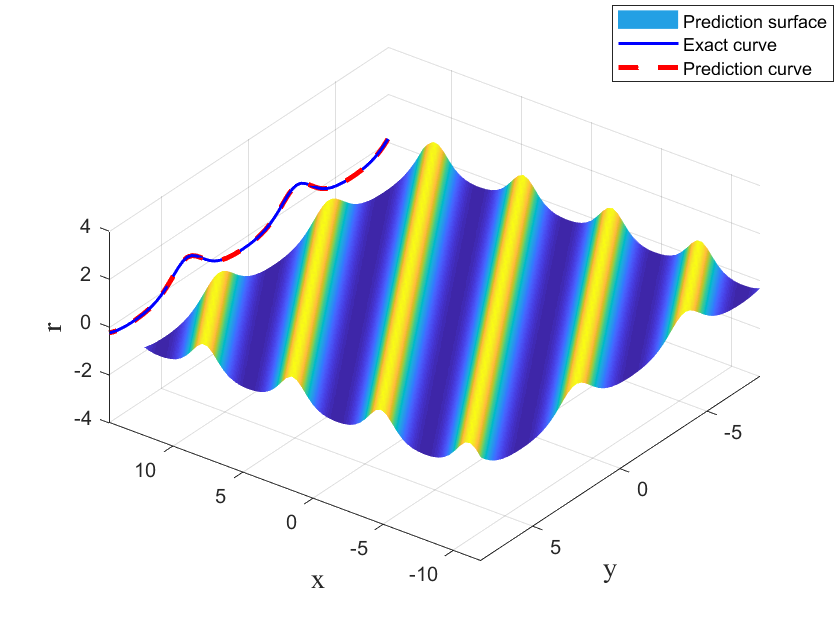}}}
~~~~
{\rotatebox{0}{\includegraphics[width=4.0cm,height=4.0cm,angle=0]{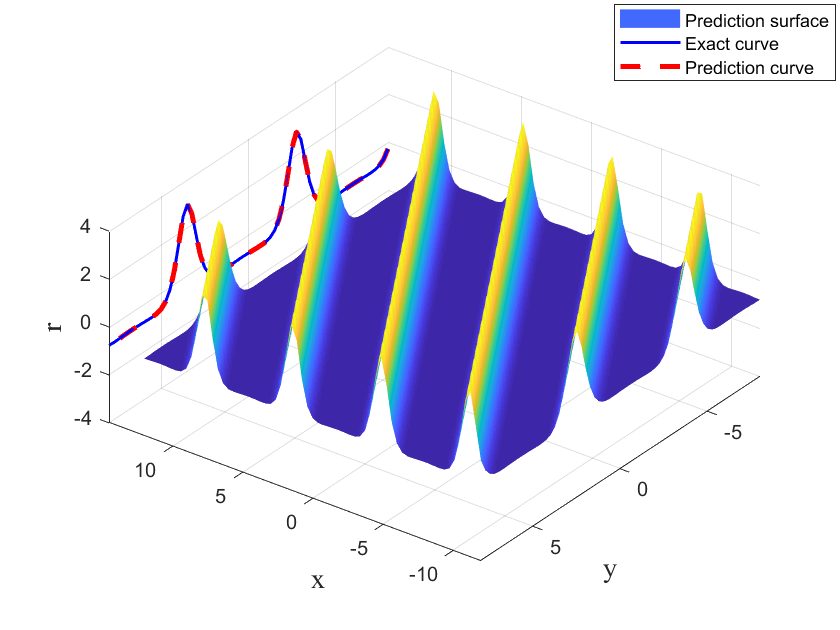}}}
~~~~
{\rotatebox{0}{\includegraphics[width=4.0cm,height=4.0cm,angle=0]{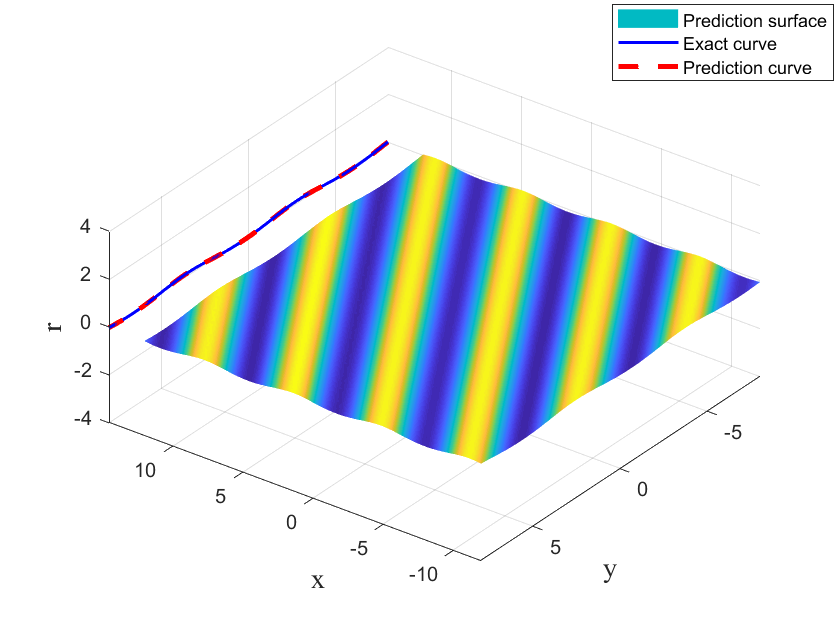}}}

$\quad\qquad\quad\quad\textbf{(d)}\quad\qquad\qquad\qquad\qquad\quad\quad\textbf{(e)}\qquad\qquad\qquad\quad\qquad\quad\textbf{(f)}$\\
\noindent { \small \textbf{Figure 4.} (Color online)  Solving the data-driven breather wave solution $q(x,y,t)$ (see $\textbf{(a),(b),(c)}$), $r(x,y,t)$ (see $\textbf{(d),(e),(f)}$) in the $(x, y)$ plane for nonlocal (2+1)-dimensional NLS equation \eqref{15} via PTS-PINN:
$\textbf{(a),(d)}$ The three-dimensional plot at $t=-1$;
$\textbf{(b),(e)}$ The three-dimensional plot at $t=0$;
$\textbf{(c)(f)}$ The three-dimensional plot at $t=4$.}\\

~~~~~~~~~~~~{\rotatebox{0}{\includegraphics[width=5cm,height=5cm,angle=0]{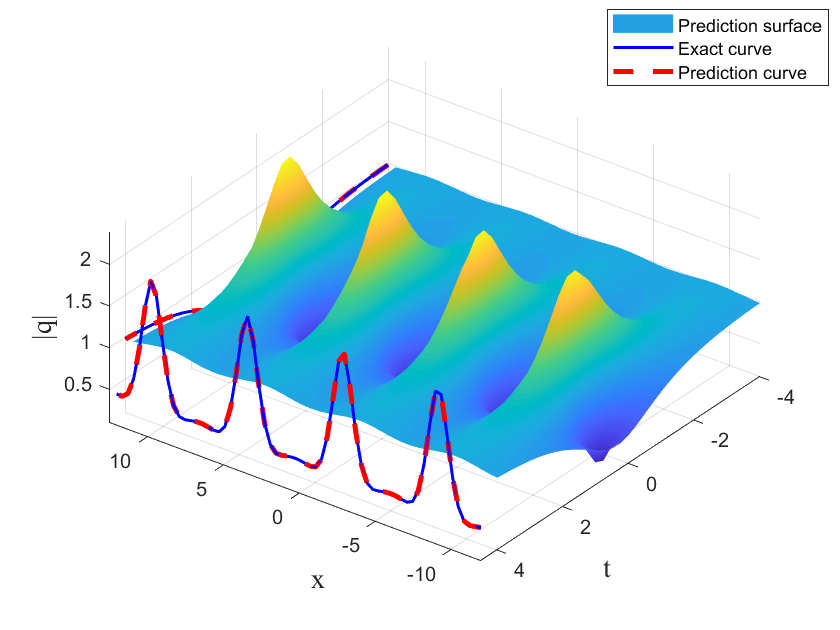}}}~~~~~~
{\rotatebox{0}{\includegraphics[width=5cm,height=5cm,angle=0]{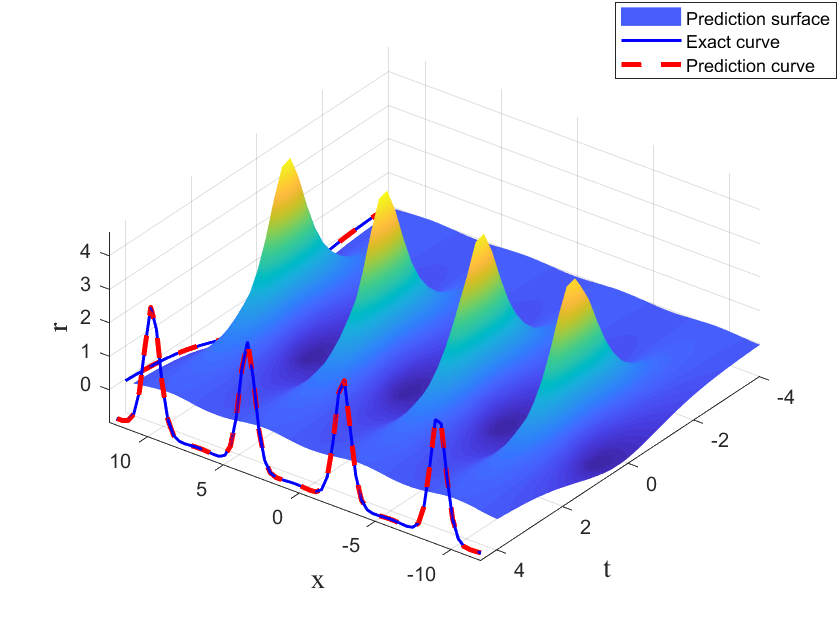}}}\\
\noindent { \small \textbf{Figure 5.} (Color online) Solving the data-driven breather wave solution $q(x,y,t), r(x,y,t)$ in the $(x, t)$ plane at $y=0$ for nonlocal (2+1)-dimensional NLS equation \eqref{15} via PTS-PINN.}\\

\begin{table}[htbp]
  \caption{nonlocal (2+1)-dimensional NLS equation: $\mathbb{L}_{2}$ relative errors of PTS-PINN for different number of hidden layers and neurons, $N_{IB}=3000, N_{f}=20000$}
  \label{Tab5}
  \centering
  \resizebox{\textwidth}{!}{
  \begin{tabular}{l|cccc}
  \toprule
  \diagbox{\textbf{Layers}}{\textbf{Neurons}} & 40 & 50 & 60\\
  \hline
  5   & q: 5.125868e-04 r: 2.109223e-03 & q: 4.717451e-02 r: 1.583170e-01 & q: 4.923205e-04 r: 2.103156e-03\\
  6   & q: 6.421177e-04 r: 2.315614e-03 & q: 2.957248e-04 r: 1.304059e-03 & q: 1.957664e-01 r: 8.207638e-01\\
  7   & q: 1.726941e-04 r: 6.785540e-04 & q: 3.206385e-04 r: 1.352906e-03 & q: 2.510787e-04 r: 8.843592e-04\\
  \bottomrule
  \end{tabular}}
\end{table}

\begin{table}[htbp]
  \caption{nonlocal (2+1)-dimensional NLS equation: $\mathbb{L}_{2}$ relative errors of PTS-PINN for different number of  IB points and collocation points, 5 layer neural network and 40 neurons per layer}
  \label{Tab6}
  \centering
  \resizebox{\textwidth}{!}{
  \begin{tabular}{l|cccc}
  \toprule
  \diagbox{\textbf{IB points}}{\textbf{Collocation points}} & 20000 & 25000 & 30000\\
  \hline
  2000   & q: 5.498626e-04 r: 2.165285e-03 & q: 5.001512e-04 r: 1.952543e-03 & q: 9.084020e-04 r: 3.490624e-03\\
  3000   & q: 5.125868e-04 r: 2.109223e-03 & q: 3.834830e-04 r: 1.511158e-03 & q: 2.048385e-01 r: 8.356843e-01\\
  4000   & q: 3.907333e-04 r: 1.460875e-03 & q: 2.177310e-04 r: 7.065225e-04 & q: 4.681463e-04 r: 1.923369e-03\\
  \bottomrule
  \end{tabular}}
\end{table}

\section{Inverse problem of nonlocal equation via PTS-PINN}
In this section, our focus is on investigating the inverse problem associated with nonlocal equations using the PTS-PINN approach. Specifically, we delve into the nonlocal (2+1)-dimensional NLS equation and the nonlocal three wave interaction systems.
The nonlocal (2+1)-dimensional NLS equation is considered in parameter form
\begin{equation}
\begin{aligned}\label{17}
&iq_{t}+aq_{xy}+qr=0,\quad x\in [-10,10],\ y\in [-10,10],\ t \in [-10,10],\\
&r_{y}=b[q(x,y,t)q(-x,-y,t)^{\ast}]_{x}.
\end{aligned}
\end{equation}
As an illustrative example, we set $a=1$ and $b=1$ to showcase the efficacy of PTS-PINN in learning the two parameters
$a$ and $b$ in Eq.\eqref{17}, along with the corresponding solution, given by \cite{Peng-MMAS52}
\begin{align}\label{18}
q(x,y,t)=1-\frac{12it+18}{(x+3y)^{2}+2t^{2}+4.5},\quad r(x,y,t)=\frac{-12(x+3y)^{2}+24t^{2}+54}{((x+3y)^{2}+2t^{2}+4.5)^{2}}.
\end{align}

The nonlocal three wave interaction systems, presented in parameter form as:
\begin{equation}
\begin{aligned}\label{19}
&q_{1t}+aq_{1x}-q_{2}(-x,-t)q_{3}(-x,-t)=0,\\
&q_{2t}+bq_{2x}-q_{1}(-x,-t)q_{3}(-x,-t)=0,\\
&q_{3t}+cq_{3x}+q_{1}(-x,-t)q_{2}(-x,-t)=0,
\end{aligned}
\end{equation}
where $x\in [-5,5]$ and $t \in [-5,5]$, serves as the focus of our investigation. For the purpose of illustrating the effectiveness of PTS-PINN, we select $a=1, b=2$ and $c=3$ as exemplary parameter values in Equation \eqref{19}. Our aim is to demonstrate the capability of PTS-PINN in learning these parameters and obtaining the corresponding solution, given by \cite{Ab-siam}
\begin{equation}
\begin{aligned}\label{20}
&q_{1}(x,t)=-\frac{2\sqrt{6}e^{t-x}(-1+e^{12t-4x})}{3e^{14t-6x}+e^{12t-4x}+e^{2t-2x}+3},\\
&q_{2}(x,t)=-\frac{16e^{7t-3x}}{3e^{14t-6x}+e^{12t-4x}+e^{2t-2x}+3},\\
&q_{3}(x,t)=\frac{4\sqrt{6}e^{-6t+2x}(1+e^{-2t+2x})}{3e^{-14t+6x}+e^{-12t+4x}+e^{-2t+2x}+3}.
\end{aligned}
\end{equation}

\subsection{Numerical experiment of nonlocal (2+1)-dimensional NLS}

By employing Latin Hypercube Sampling, we can generate a training data set by randomly selecting $N_{l}=5000$ as the IB value data and $N_{f}=5000$ as collocation points. The data set is constructed using the exact rogue wave solution \eqref{18}, and the spatial coordinates $(x, y, t)$ are discretized into 64 points within the range $[-10, 10]\times [-10, 10]\times [-10, 10]$, respectively.
In the PTS-PINN scheme, utilizing the acquired training data set, a 6-hidden-layer neural network with 40 neurons per layer is employed to predict the solutions $q(x, y, t), r(x, y, t)$, as well as the unknown parameters $a, b$. In the absence of noise, the data-driven solutions $q(x,y,t), r(x,y,t)$ display an $\mathbb{L}_{2}$-norm error of 2.298788e-04 and 1.894221e-03 when compared to the exact solutions, respectively.
 The three-dimensional contour diagrams at $(x,y)$ plane, two-dimensional cross-section diagram
of the data-driven rogue wave solutions are clearly shown in  Fig. 6, and the three-dimensional contour diagrams at $(x,t)$  plane are also presented at Fig. 7. Table 7 presents the accurate parameters of the nonlocal (2+1)-dimensional NLS equation alongside the learned parameters $a$ and $b$ using PTS-PINN. Notably, PTS-PINN demonstrates a remarkable ability to correctly identify unknown parameters with very high accuracy, particularly when the training data is devoid of noise. Additionally, even in the presence of 0.05 noise and 0.1 noise, the errors associated with parameters $a$ and $b$ remain acceptable, suggesting the robustness of the predictions. It is worth noting that noise has a negative impact on parameter error values.

{\rotatebox{0}{\includegraphics[width=4.0cm,height=4.0cm,angle=0]{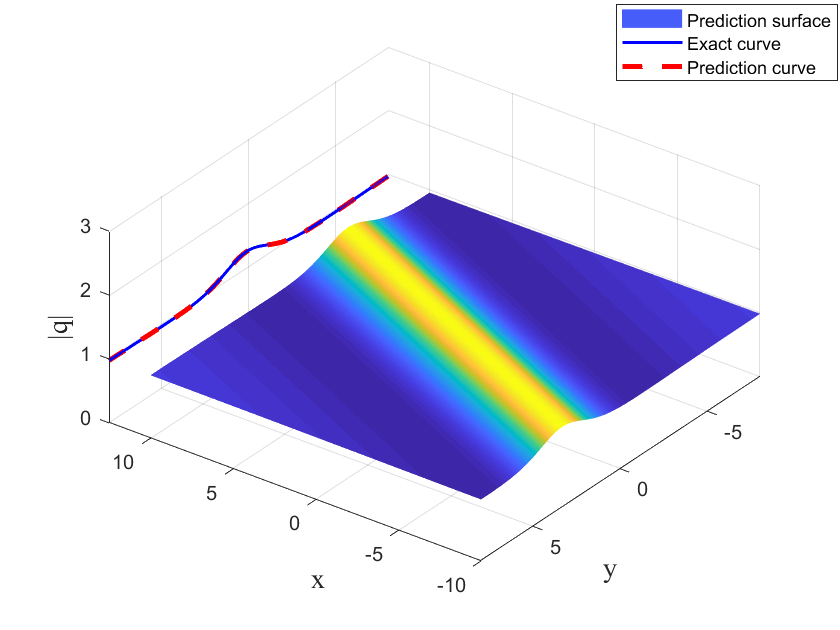}}}
~~~~
{\rotatebox{0}{\includegraphics[width=4.0cm,height=4.0cm,angle=0]{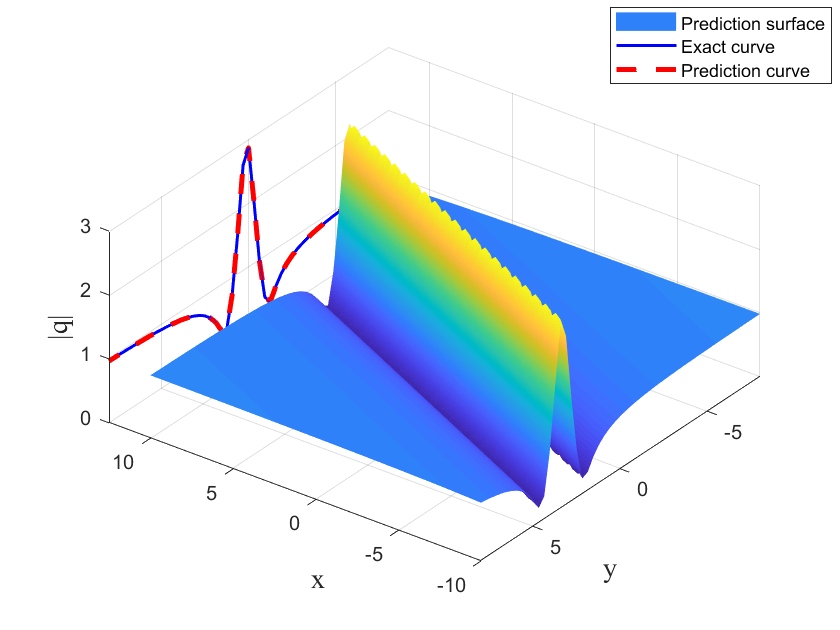}}}
~~~~
{\rotatebox{0}{\includegraphics[width=4.0cm,height=4.0cm,angle=0]{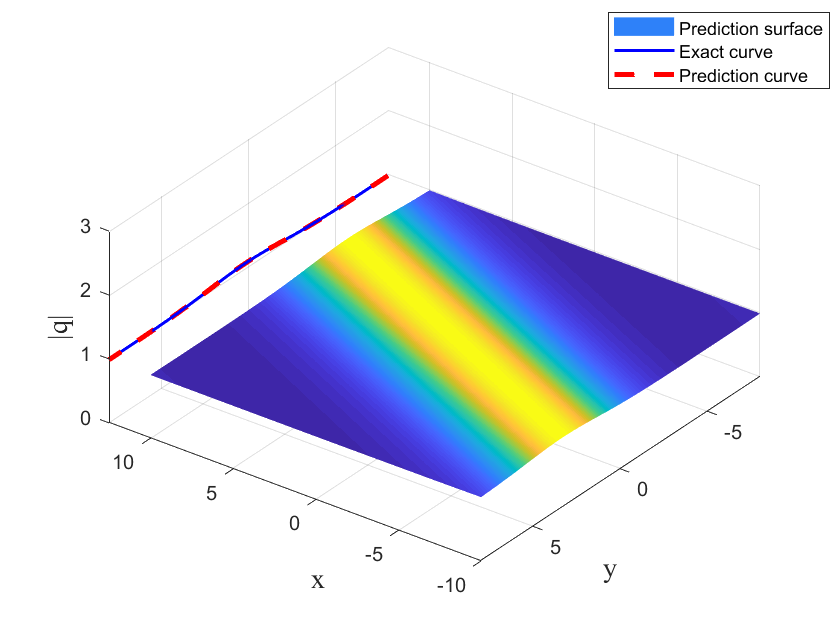}}}

$\quad\qquad\quad\quad\textbf{(a)}\quad\qquad\qquad\qquad\qquad\quad\quad\textbf{(b)}\qquad\qquad\qquad\quad\qquad\quad\textbf{(c)}$

{\rotatebox{0}{\includegraphics[width=4.0cm,height=4.0cm,angle=0]{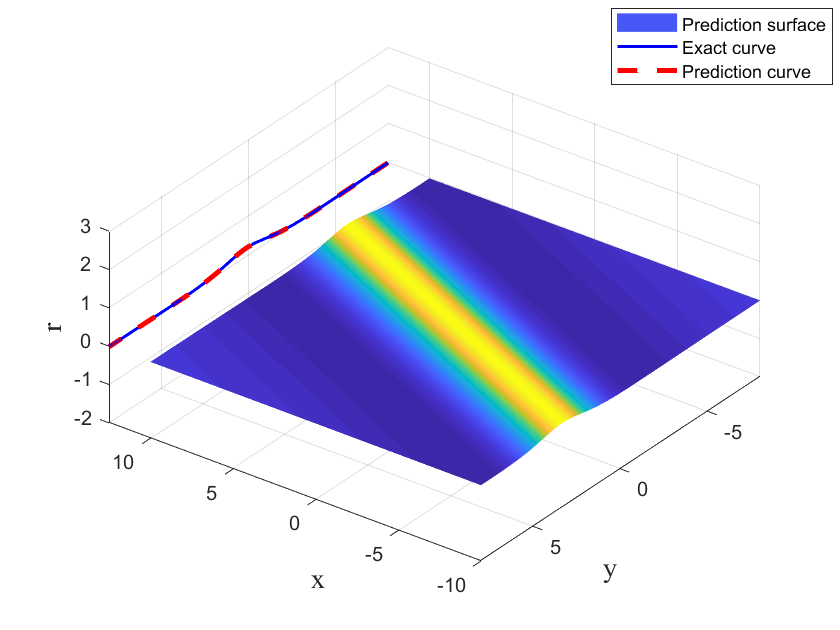}}}
~~~~
{\rotatebox{0}{\includegraphics[width=4.0cm,height=4.0cm,angle=0]{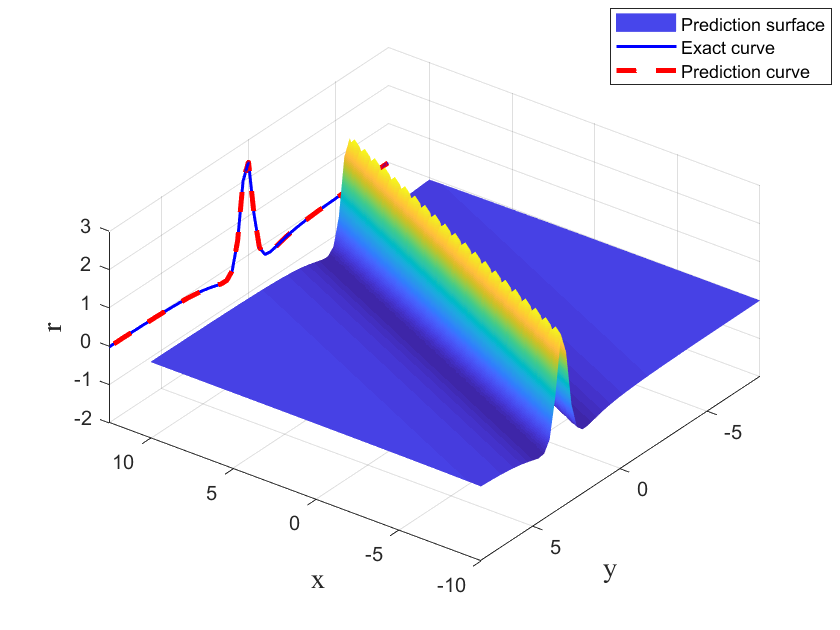}}}
~~~~
{\rotatebox{0}{\includegraphics[width=4.0cm,height=4.0cm,angle=0]{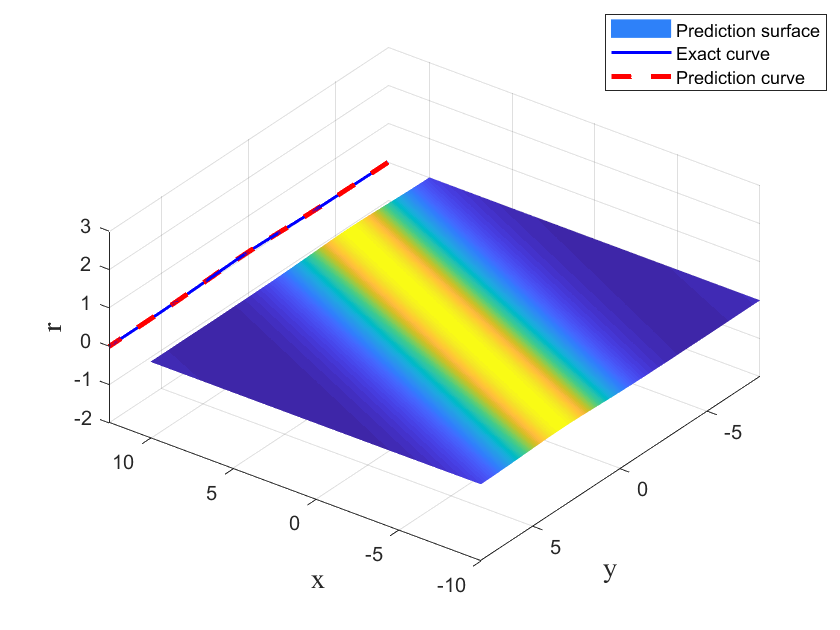}}}

$\quad\qquad\quad\quad\textbf{(d)}\quad\qquad\qquad\qquad\qquad\quad\quad\textbf{(e)}\qquad\qquad\qquad\quad\qquad\quad\textbf{(f)}$\\
\noindent { \small \textbf{Figure 6.} (Color online)  Solving the data-driven rogue wave solution $q(x,y,t)$ (see $\textbf{(a),(b),(c)}$), $r(x,y,t)$ (see $\textbf{(d),(e),(f)}$) in the $(x, y)$ plane for nonlocal (2+1)-dimensional NLS equation \eqref{17} via PTS-PINN:
$\textbf{(a),(d)}$ The three-dimensional plot at $t=-5$;
$\textbf{(b),(e)}$ The three-dimensional plot at $t=0$;
$\textbf{(c)(f)}$ The three-dimensional plot at $t=10$.}\\

~~~~~~~~~~~~{\rotatebox{0}{\includegraphics[width=5cm,height=5cm,angle=0]{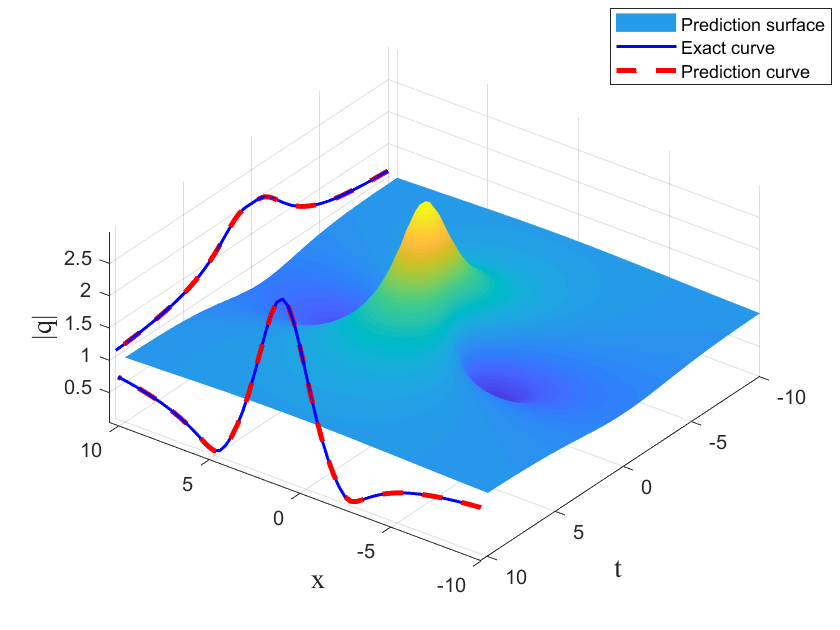}}}~~~~~~
{\rotatebox{0}{\includegraphics[width=5cm,height=5cm,angle=0]{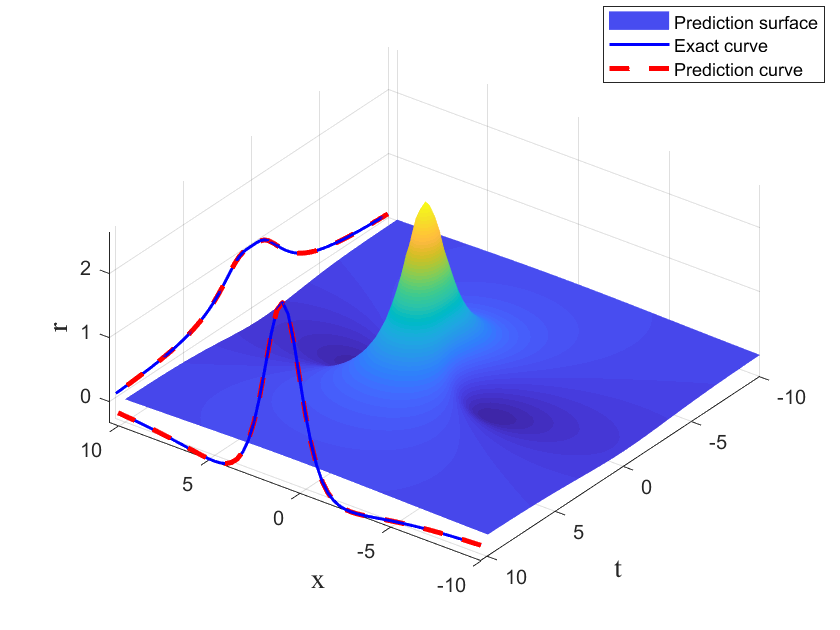}}}\\
\noindent { \small \textbf{Figure 7.} (Color online) Solving the data-driven rogue wave solutions $q(x,y,t)$, $r(x,y,t)$ in the $(x, t)$ plane at $y=0$ for nonlocal (2+1)-dimensional NLS equation \eqref{17} via PTS-PINN.}\\

\begin{table}[htbp]
  \caption{Data-driven parameter discovery of $a, b$}
  \label{Tab7}
  \centering
  \resizebox{\textwidth}{!}{
  \begin{tabular}{l|cccc}
  \toprule
  \diagbox{\textbf{Noise}}{\textbf{Parameter}} & $a$ & error of $a$ & $b$ & error of $b$\\
  \hline
  Correct parameter   & 1& 0& 1& 0\\
  Without noise   &0.9994288 & 0.05712\% & 1.000033 & 0.00328\%\\
  With a 0.05  noise   &0.9990587&0.09413\% & 0.9987529&0.12471\%\\
  With a 0.1  noise   &0.9983717&0.16283\% & 0.9962330&0.37670\%\\
  \bottomrule
  \end{tabular}}
\end{table}

In the context of the inverse problem, we scrutinize the evolution of unknown parameters and the loss function with varying numbers of iterations across different noise levels. Figs. 8(a) and (b) depict the alterations in unknown parameters with iteration under different noise conditions. Remarkably, we observe that, even in the presence of noise, PTS-PINN enables effective learning of unknown parameters after approximately 300 iterations. Fig. 8(c) illustrates the fluctuations in the loss function under different noise levels as the number of iterations increases. The findings indicate that as the noise level escalates, the convergence effect gradually diminishes. This underscores the impact of noise on the convergence behavior of the learning process.
\\
{\rotatebox{0}{\includegraphics[width=4.6cm,height=4.0cm,angle=0]{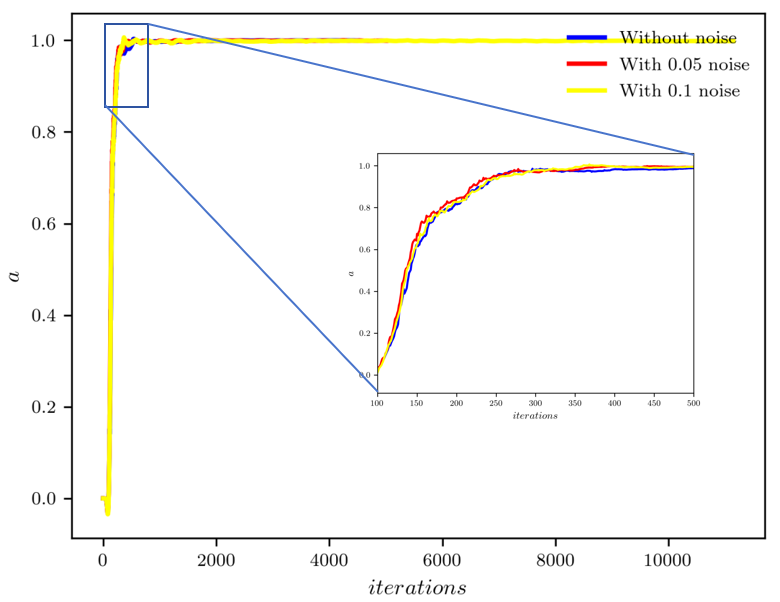}}}
~~~~
{\rotatebox{0}{\includegraphics[width=4.6cm,height=4.0cm,angle=0]{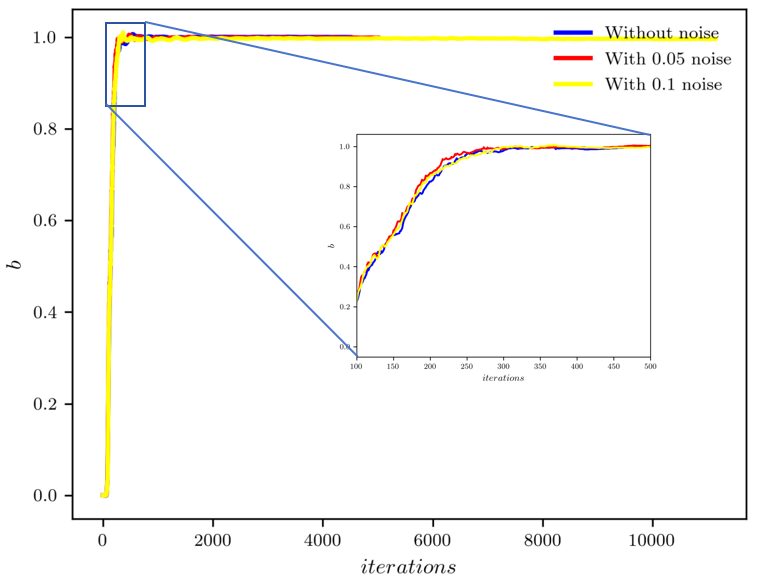}}}
~~~~
{\rotatebox{0}{\includegraphics[width=4.6cm,height=4.0cm,angle=0]{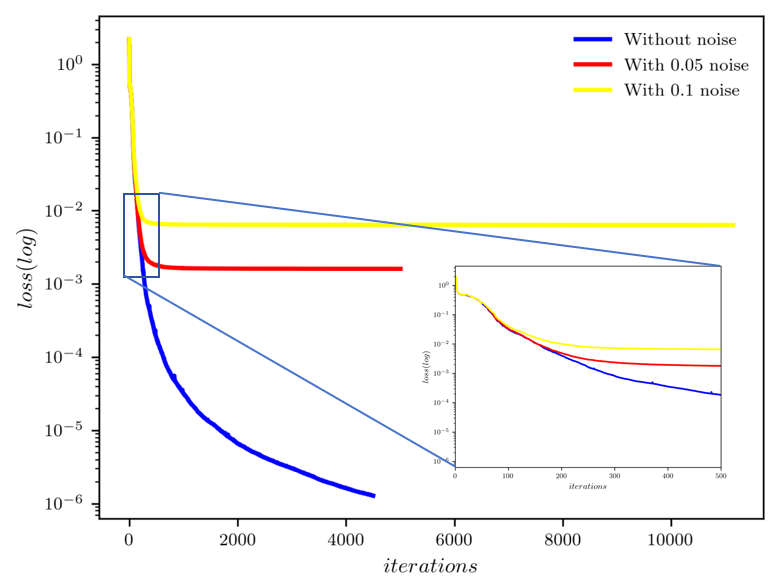}}}\\
$~~~~~~~~~~~~~~~(\textbf{a})~~~~~~~~~~~~~~~~
~~~~~~~~~~~~~~~~~~~~~~~~~(\textbf{b})~~~~~~~
~~~~~~~~~~~~~~~~~~~~~~~~~~~~(\textbf{c})$\\
\noindent { \small \textbf{Figure 8.} (Color online)$\textbf{(a, b)}$ the variation of unknown parameters $a, b$ and $\textbf{(c)}$  the variation of loss function with the different noise.}\\

\subsection{Numerical experiment of nonlocal three wave interaction systems}
Randomly selecting $N_{l}=1000$ as the IB data and $N_{f}=1000$ as the collocation points, with the aid of the exact soliton solution \eqref{20}, and considering $(x, t) \in [-5, 5]\times [-5, 5]$ divided into 512 points and 400 points, respectively, a 9-hidden-layer deep PTS-PINN with 40 neurons per layer is employed. This architecture is utilized to predict the solutions $q_{1}(x,t),q_{2}(x,t),q_{3}(x,t)$, as well as the unknown parameters $a, b, c$. The three-dimensional graphs of the predicted solution are illustrated in Fig. 9, while Fig. 10 displays the absolute error surface of $q_{1}(x,t),q_{2}(x,t),q_{3}(x,t)$. Notably, the absolute error is approximately e-3, indicating a high level of accuracy in the predictions.
Table 8 presents the accurate parameters of the nonlocal three wave interaction system alongside the parameters $a, b$ and $c$ obtained through the learning process using PTS-PINN. Similar to the observations in Table 8, it is evident that the impact of noise on parameter error values is noticeable. However, PTS-PINN continue to exhibit a high level of accuracy even in the presence of noise, emphasizing the robustness of the approach.

{\rotatebox{0}{\includegraphics[width=4cm,height=4cm,angle=0]{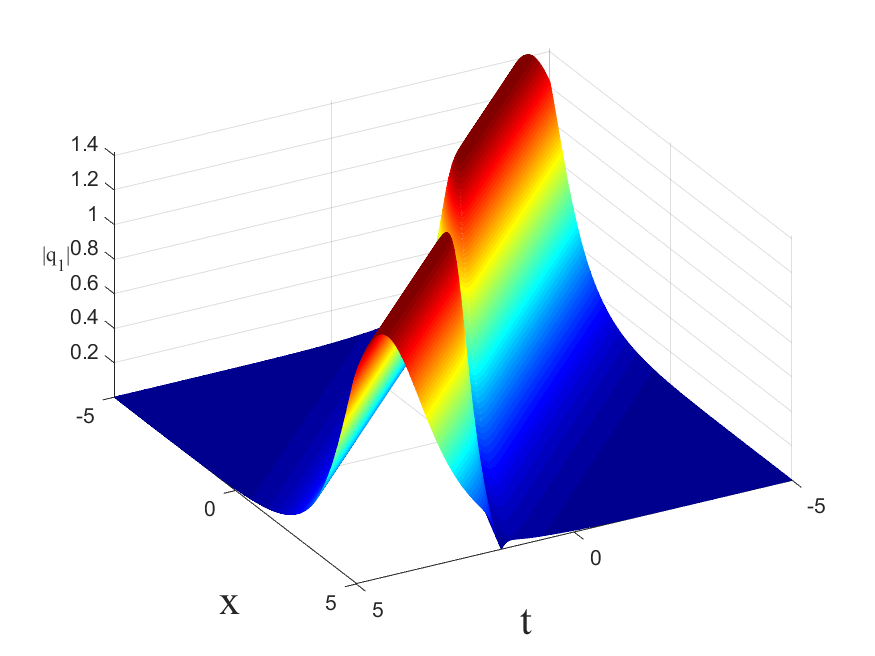}}}
~~~~
{\rotatebox{0}{\includegraphics[width=4cm,height=4cm,angle=0]{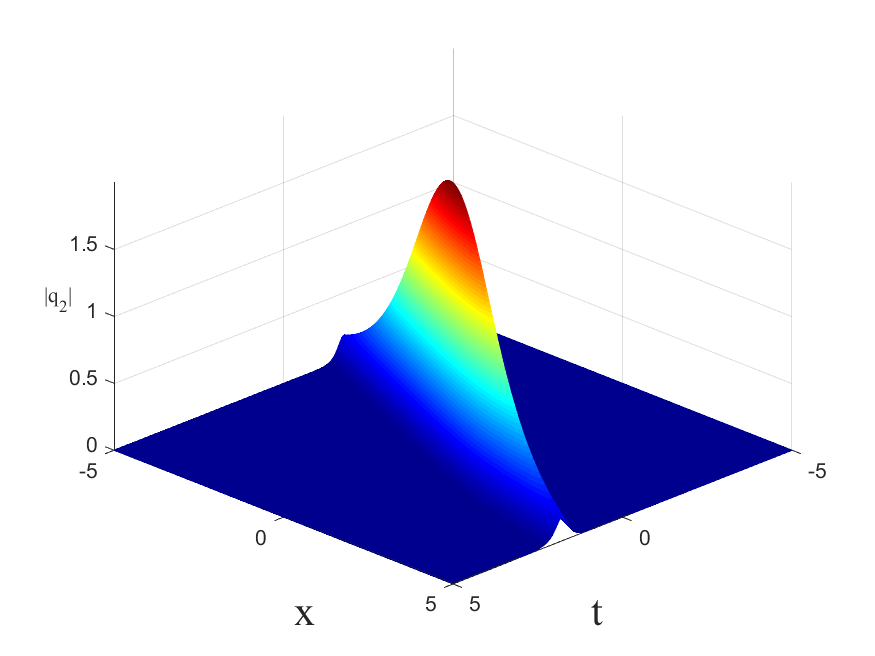}}}
~~~~
{\rotatebox{0}{\includegraphics[width=4cm,height=4cm,angle=0]{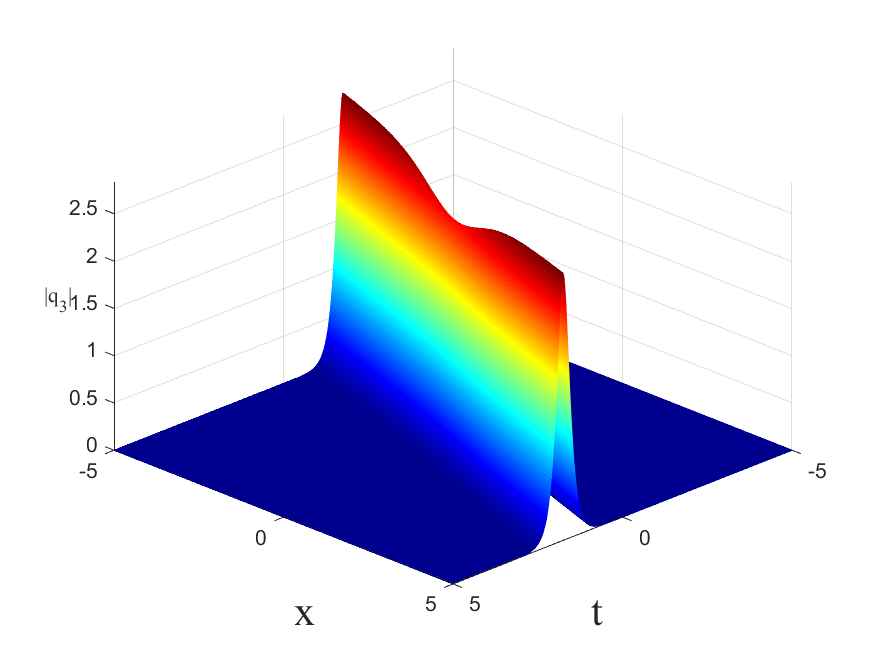}}}\\
\noindent { \small \textbf{Figure 9.} (Color online) Solving the data-driven soliton wave solution $q_{1}(x,t), q_{2}(x,t),q_{3}(x,t)$ for nonlocal three wave interaction systems  \eqref{19} via PT-PINN.}\\

{\rotatebox{0}{\includegraphics[width=4.0cm,height=4.0cm,angle=0]{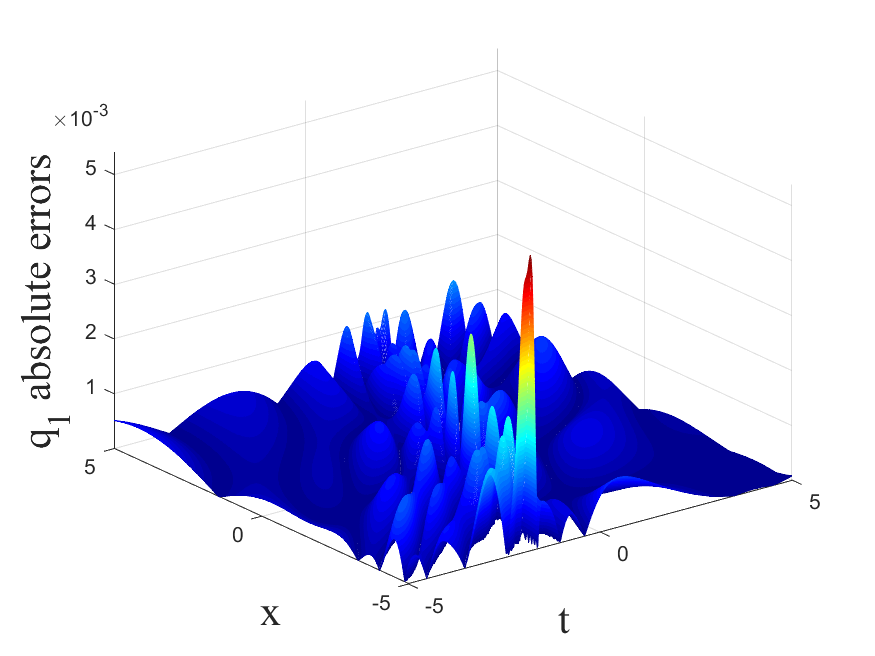}}}
~~~~
{\rotatebox{0}{\includegraphics[width=4.0cm,height=4.0cm,angle=0]{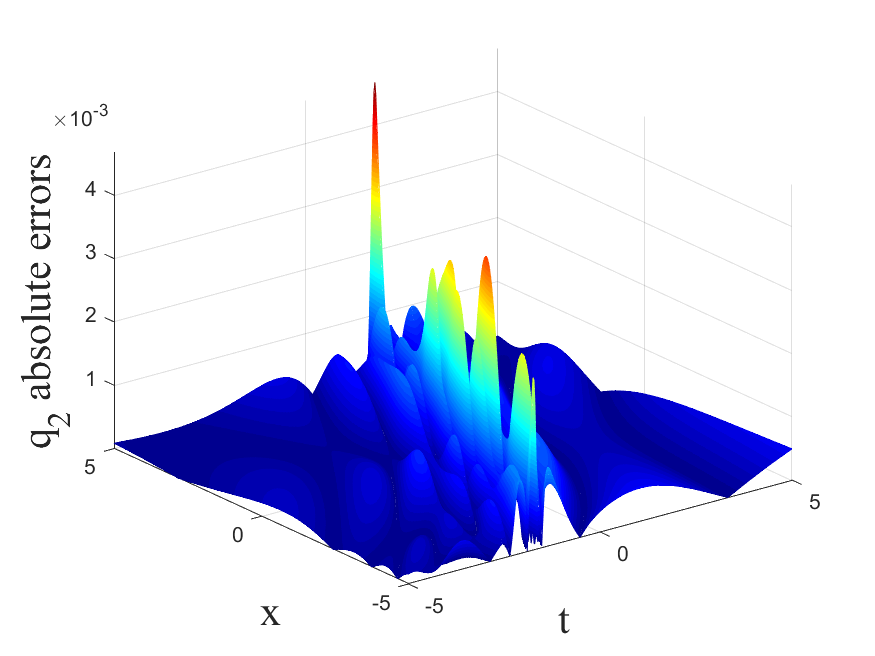}}}
~~~~
{\rotatebox{0}{\includegraphics[width=4.0cm,height=4.0cm,angle=0]{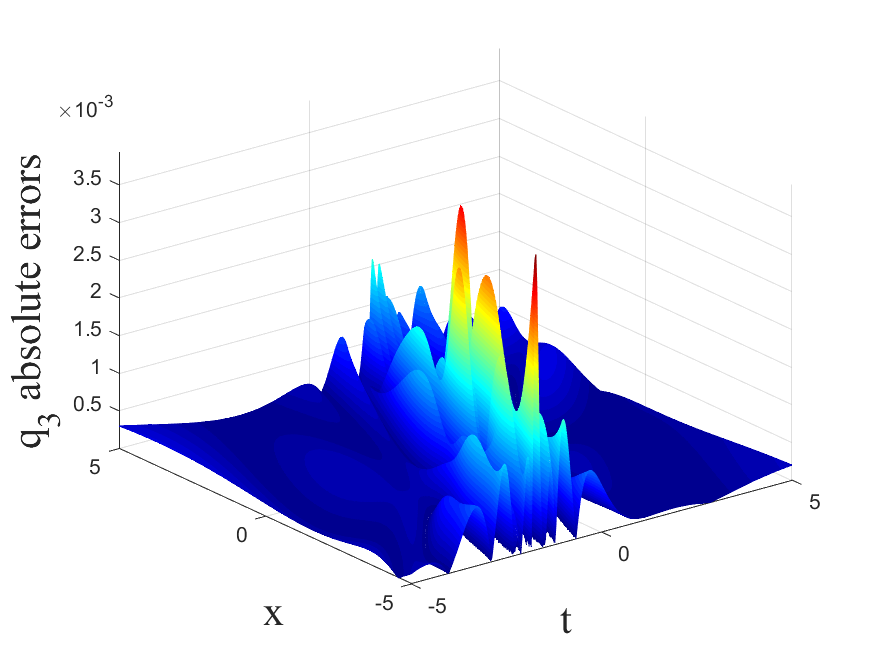}}}\\
\noindent { \small \textbf{Figure 10.} (Color online)Nonlocal three wave interaction systems: Error Dynamics for $q_{1}, q_{2}, q_{3}$.}\\

\begin{table}[htbp]
  \caption{Data-driven parameter discovery of $a, b, c$}
  \label{Tab8}
  \centering
  \resizebox{\textwidth}{!}{
  \begin{tabular}{l|cccccc}
  \toprule
  \diagbox{\textbf{Noise}}{\textbf{Parameter}} & $a$ & error of $a$ & $b$ & error of $b$ & $c$ & error of $c$\\
  \hline
  Correct parameter   & 1& 0& 2& 0 & 3& 0\\
  Without noise   &1.000113 & 0.01125\% & 1.999889 & 0.00554\% & 3.000028 & 0.00093\%\\
  With a 0.05  noise   &0.9982466&0.17534\% & 1.997929&0.10355\% & 3.000673 & 0.02243\%\\
  With a 0.1  noise   &1.003294&0.32938\% & 1.979039&1.04805\% & 3.048382 & 1.61273\%\\
  \bottomrule
  \end{tabular}}
\end{table}

Figs. 11(a), (b), and (c) analyze the changes of unknown parameters and loss functions with the number of iterations under different noise levels. From these figures, we observe that parameter $a$ can be learned after approximately 200 iterations, while parameter $b$ requires approximately 600 iterations, and parameter $c$ requires approximately 100 iterations. Similar to Fig. 8, for the nonlocal three wave interaction systems, PTS-PINN can find parameters well even in the presence of noise. However, noise also has a negative impact on the convergence effect of the loss function.
\\
{\rotatebox{0}{\includegraphics[width=3.6cm,height=3.0cm,angle=0]{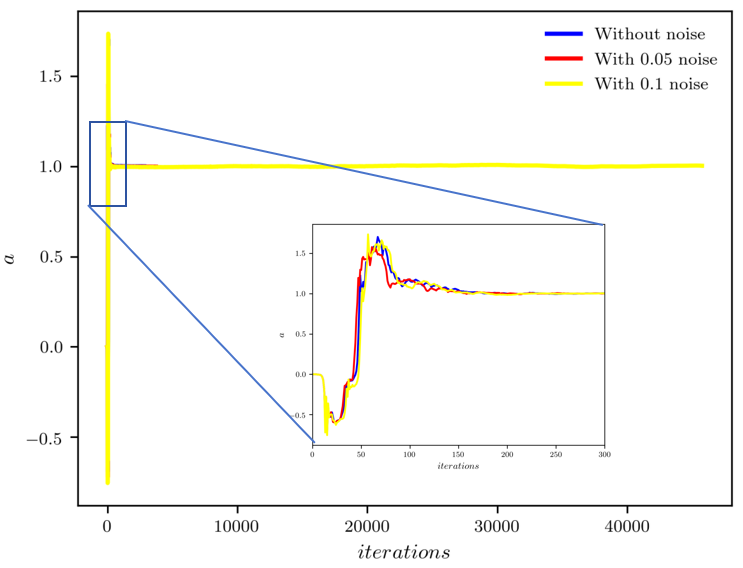}}}~
{\rotatebox{0}{\includegraphics[width=3.6cm,height=3.0cm,angle=0]{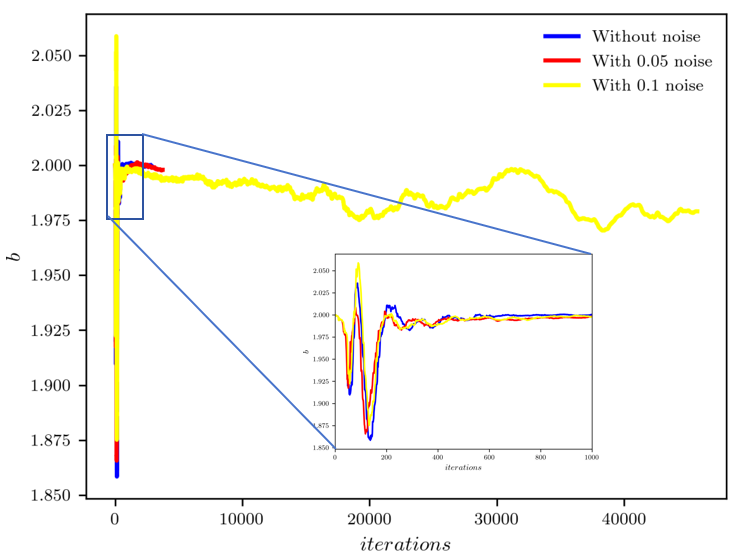}}}~
{\rotatebox{0}{\includegraphics[width=3.6cm,height=3.0cm,angle=0]{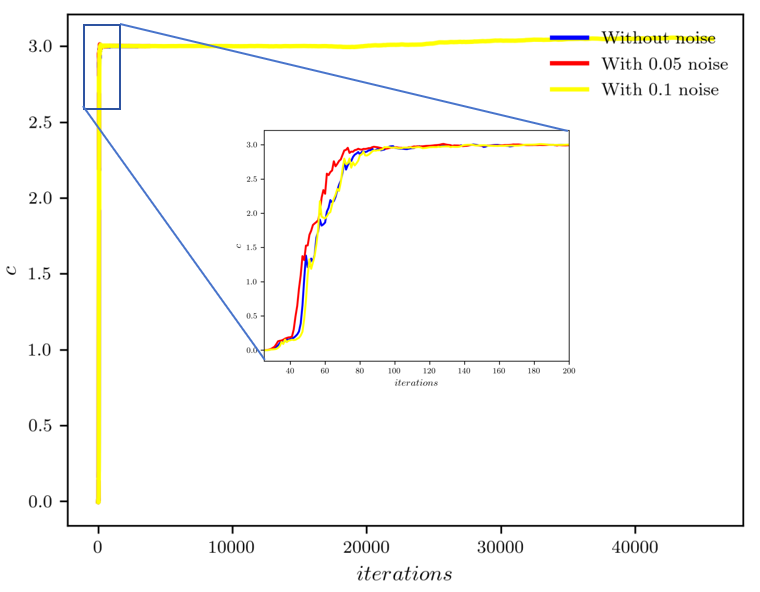}}}~
{\rotatebox{0}{\includegraphics[width=3.6cm,height=3.0cm,angle=0]{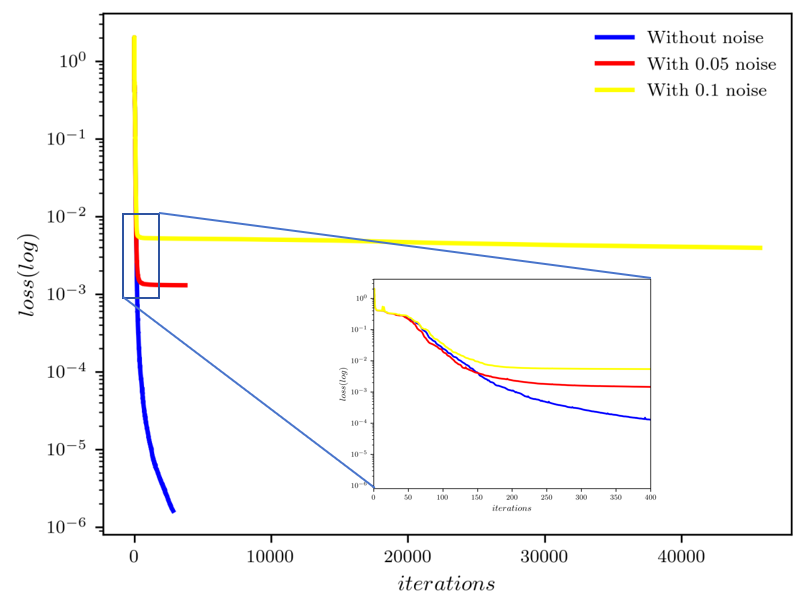}}}\\
$~~~~~~~~~~~~~~~(\textbf{a})~~~~
~~~~~~~~~~~~~~~~~~~~~(\textbf{b})
~~~~~~~~~~~~~~~~~~~~~~~~(\textbf{c})~~
~~~~~~~~~~~~~~~~~~~~~~(\textbf{d})$\\
\noindent { \small \textbf{Figure 11.} (Color online)$\textbf{(a, b, c)}$ the variation of unknown parameters $a, b, c$ and $\textbf{(d)}$  the variation of loss function with the different noise.}\\

\section{Conclusion}

In this paper, we aim to enhance the effectiveness of the PINN algorithm in solving integrable nonlocal models by incorporating the $PT$-symmetry property of nonlocal equations into the PINN loss function. Through a systematic study of the performance of PTS-PINN with varying numbers of IB points, collocation points, hidden layers, and neurons per layer, we verify that PTS-PINN algorithm is highly effective. In particular, we aim to highlight the exceptional ability of the PTS-PINN algorithm in learning rogue waves associated with the nonlocal equation even at large spatiotemporal scales.  Furthermore, we investigate the performance of PTS-PINN in solving the nonlocal derivative NLS equation. For learning periodic waves, the precision of the PTS-PINN algorithm reaches the order of e-04.
We extend our challenge to accurately solve the nonlocal (2+1)-dimensional NLS equation with PTS-PINN, and the data-driven breather wave solution is perfectly learned. Detailed numerical experimental results on the aforementioned nonlocal PDEs are presented to illustrate the effectiveness of the PTS-PINN method. Additionally, we employ PTS-PINN to determine the parameters involved in the nonlocal (2+1)-dimensional NLS equation and nonlocal three wave interaction systems. Experimental results consistently showcase the excellent performance of the PTS-PINN algorithm in solving the inverse problems of nonlocal equations. We believe that our results will contribute to the understanding and resolution of other nonlocal integrable systems.

\section*{Acknowledgements}
\hspace{0.3cm}
This work was supported by the National Natural Science Foundation of China (No. 12175069 and No. 12235007), Science and Technology Commission of Shanghai Municipality (No. 21JC1402500 and No. 22DZ2229014) and Natural Science Foundation of Shanghai (No. 23ZR1418100).

\end{document}